\documentclass[onecolumn]{aastex631}
\UseRawInputEncoding
\usepackage{mathtools}
\usepackage{xcolor}
\usepackage{natbib}
\definecolor{BLUE}{rgb}{0.2,0.2,1}

\graphicspath{{./}{figures/}}

\usepackage{mathrsfs}
\usepackage{amssymb}
\usepackage{amsmath}
\usepackage{graphicx}
\usepackage[normalem]{ulem}
\usepackage{bm}
\usepackage{longtable}
\usepackage{slashed}
\renewcommand\theequation{\arabic{equation}}
\usepackage[T1]{fontenc}
\usepackage{times}
\usepackage{fouriernc}
\usepackage{array}

\usepackage{hyperref}
\hypersetup{colorlinks,linkcolor={cyan},citecolor={magenta},urlcolor={magenta}}  

\renewcommand\sout{\bgroup \color{red} \ULdepth=-.5ex \ULset}

\renewcommand{\v}[1]{\textbf{#1}}
\renewcommand{\rm}[1]{\textrm{#1}}
\renewcommand{\d}{\mathrm{d}}

\renewcommand{\baselinestretch}{0.98}

\makeatletter
\let\frontmatter@title@above=\relax
\makeatother

\renewcommand\thesection{\arabic{section}}

\let\oldbibliography\thebibliography
\renewcommand{\thebibliography}[1]{
  \oldbibliography{#1}
  \setlength{\itemsep}{1pt}
  \setlength{\baselineskip}{10.pt}
  \setlength{\lineskiplimit}{-\maxdimen}
}
\usepackage{jabbrv}

\usepackage{subfigure,dcolumn,jabbrv}
\usepackage{graphicx}
\usepackage{dcolumn}
\usepackage{bm}
\usepackage{CJK}
\usepackage{url}
\usepackage{color}
\usepackage{booktabs}
\usepackage{natbib}
\usepackage{hyperref}
\usepackage{cleveref}
\renewcommand{\thefootnote}{$\star$}

\begin{document}

\title{Bayesian Inference of Hybrid Star Properties\\ from Future High-Precision Measurements of Their Radii}

\author{Bao-An Li\footnote{Corresponding Author:Bao-An.Li@etamu.edu}}
\affiliation{Department of Physics and Astronomy, East Texas A$\&$M University, Commerce, TX 75429-3011, USA}
\author{Xavier Grundler}
\affiliation{Department of Physics and Astronomy, East Texas A$\&$M University, Commerce, TX 75429-3011, USA}
\author{Wen-Jie Xie}
\affiliation{Department of Physics and Astronomy, East Texas A$\&$M University, Commerce, TX 75429-3011, USA}
\affiliation{Department of Physics, Yuncheng University, Yuncheng 044000, China}
\affiliation{Guangxi Key Laboratory of Nuclear Physics and Nuclear Technology, Guangxi Normal University, Guilin 541004, China}
\author{Nai-Bo Zhang}
\affiliation{Department of Physics and Astronomy, East Texas A$\&$M University, Commerce, TX 75429-3011, USA}
\affiliation{School of Physics, Southeast University, Nanjing 211189, China}
\date{\today}

\fontdimen2\font=2.pt

\begin{abstract}
Future high-precision X-ray and gravitational-wave observations of neutron stars (NSs) are expected to constrain NS radii with uncertainties as small as $\sigma \simeq 0.1$~km. Such unprecedented precision offers a unique opportunity to extract new information about the nature and equation of state (EOS) of supradense matter in NS cores. Using mock radius data with uncertainties ranging from $\sigma = 1.0$ to $0.1$~km, together with a flexible meta-model NS EOS that allows for a first-order hadron--quark phase transition, we perform a Bayesian statistical analysis to assess the impact of radius measurements on EOS constraints. We find that high-precision radius measurements, particularly for massive NSs, significantly tighten constraints on the hadron--quark transition density $\rho_t$, the quark matter mass fraction in NS cores, and several parameters characterizing the EOS of supranuclear hadronic matter, although the degree of improvement depends on the assumed prior range of $\rho_t$. In contrast, even with the highest precision considered, NS radii---including those of massive stars---remain largely insensitive to the stiffness of quark matter, independent of the measurement accuracy or the prior range adopted for $\rho_t$.
\end{abstract}


\section{Introduction}
Most of the existing neutron star radius data have $1\sigma$ precisions larger than 1.0 km that is about 10\% of the mean radius of canonical neutron stars with masses around 1.4 M$_{\odot}$. While these data have certainly improved our understanding about the EOS of neutron star matter, more precise radius measurements may lead to major progress and reveal possibly fundamentally new physics about supradense neutron-rich matter. For instance, to distinguish many possible EOSs predicted by nuclear many-body theories and identify twin or strange stars, it is necessary to carry out the differential mass/radius measurement ${\rm d}M/{\rm d}R$ in the mass region $(1.2-2.0)$M$_{\odot}$ \citep{Zhao20,Han22,Pro,Li:2024imk,Zhang:2024npg,Cai:2025nxn,Xavier}. For this purpose, the radius has to be measured much better than 1.0 km accuracy. Indeed, such measurements for multiple scientific goals have been proposed by using the next-generation X-ray pulse profile observatories and gravitational wave detectors. For instance, the enhanced X-ray Timing and Polarimetry mission (eXTP) \citep{eXTP:2018anb,AngLi25} to be launched around 2030 is designed to measure the radius of PSR 10740+6620 to about $\pm 6\%$ accuracy, while the Advanced Telescope for High Energy Astrophysics (NewATHENA) \citep{Ath} to be launched around 2037 can measure the radius of PSR 10740+6620 to about $\pm 3\%$ accuracy \citep{Seb}. These expected accuracies of future measurement are significantly better than the latest result from analyzing NICER observations, see, e.g., Ref. \citep{Ditt24} which concluded that the equatorial circumferential radius of PSR J0740+6620 is ${12.92}_{-1.13}^{+2.09}$ km (68\% credibility). As a reference, its asymmetric confidence boundary around the most probable radius has a lower error of about 8.8\% and an upper one of about 16.2\%. Moreover, even better precisions may be achieved with the third-generation gravitational-wave detectors \citep{Hild:2009ns,LIGOScientific:2020zkf}. For example, the Einstein Telescope \citep{Sathyaprakash:2012jk} and Cosmic Explorer \citep{Evans:2021gyd} are expected to measure the radius $R_{1.4}$ of canonical NSs to a precision better than 2.0\%, see, e.g., Refs. \citep{Chatziioannou:2021tdi,Pacilio:2021jmq,Bandopadhyay:2024zrr,Finstad:2022oni,Walker:2024loo}. More quantitatively, considering only the 75 loudest events of binary NS mergers in the one-year operation of a network consisting of one Cosmic Explorer and the Einstein Telescope, the radii of NSs in the mass range (1.00-1.97) M$_{\odot}$ are expected to be constrained to at least $\sigma\leq 0.2$ km at 90\% credibility \citep{Walker:2024loo}. 

Based on comprehensive surveys of X-ray and gravitational-wave analyses conducted since GW170817, the mean radius of a canonical NS was estimated to be $R_{1.4}=12.00\pm1.13$~km at 68\% credibility, assuming that all reported measurements are equally reliable within their quoted uncertainties \citep{LiEPJA}. Incorporating the latest constraints from nuclear experiments and theories, new astrophysical constraints and reanalyses of some earlier NS binary merger data, it was found very recently that 
$R_{1.4}=12.26^{+0.8}_{-0.91}$~km at 95\% credibility \citep{Koehn:2024set}.
Moreover, empirical evidence indicates that, within the present measurement precision of $\sigma \sim 1.0$~km, the radii of NSs with masses $1.4$, $1.6$, $1.8$, and $2.0\,M_{\odot}$ satisfy the approximate relation $R_{1.4}\simeq R_{1.6}\simeq R_{1.8}\simeq R_{2.0}$ \citep{MR-Russia,Ayriyan:2024zfw}. 
 
Relative to our current knowledge of NS radii, the planned high-precision radius measurements therefore have the potential to revolutionize our understanding of the nature of superdense matter in NS cores and to resolve many long-standing open questions. While eagerly anticipating these high-precision data, and recognizing the substantial technical challenges, systematic uncertainties, and significant investments of resources---including funding, manpower, and time---required to achieve the targeted precision, it is scientifically invaluable to assess in a timely manner the potential scientific returns of such measurements. This can be achieved by using the presently inferred mean NS radius as a reference and exploring mock data with progressively improved precisions that emulate those anticipated for forthcoming observations.

Within a Bayesian framework employing a meta-model for the NS EOS composed of neutrons, protons, electrons, and muons---the minimal hadronic $npe\mu$ NS model in $\beta$ equilibrium---we have previously investigated how prospective measurements of the mass--radius slope ${\rm d}M/{\rm d}R$, spanning negative, infinite, and positive values for NSs with masses between $1.4$ and $2.0\,M_{\odot}$ relative to the reference radius $R_{1.4}=11.9\pm1.4$~km, can inform the underlying EOS of dense NS matter \citep{xie2020bayesian}. More recently, within the same Bayesian framework and minimal NS EOS model, we examined how measurements of $R_{1.4}$ and $R_{2.0}$ with uncertainties $\sigma$ ranging from $1.0$ to $0.1$~km can progressively narrow the posterior probability distribution functions (PDFs) of EOS parameters relative to their uniform priors \citep{Li:2024imk}. We notice that some other groups, see, e.g., Ref. \citep{Mondal:2023gbf} have also obtained interesting results by using simulated high-precision data of future GW observations. 

In the present work, we extend this approach by coupling the minimal hadronic NS EOS to the Constant Sound Speed (CSS) model for quark matter \citep{Alford:2013aca,zdunik2013} via a first-order hadron--quark phase transition. We explore how increasingly precise NS radius measurements can improve constraints on the posterior PDFs of the hadron--quark transition density $\rho_t$, the quark matter fraction $F_{\rm QM}$ and its corresponding radius $R_{\rm QM}$, as well as high-density hadronic EOS parameters, particularly those governing the density dependence of the nuclear symmetry energy. Our results are expected to be valuable for both refining observational strategies and interpreting future high-precision NS radius measurements. 
Since the radii of massive NSs are almost unaffected by the existing uncertainties about NS crust, and these NSs have the best chance of hosting a quark matter core, here we use only the mock $R_{2.0}$ data.

We emphasize that, although the CSS model explicitly incorporates a first-order hadron--quark phase transition, embedding it within a Bayesian statistical framework allows us to infer directly from NS observations the relative probability of quark matter formation in NS cores, rather than assuming its presence \emph{a priori}. As we demonstrate below, the posterior probabilities for forming hybrid stars with sizable quark cores are extremely small. Consequently, in contrast to many previous studies that adopt the CSS or other quark matter models in forward modeling approaches---where the existence of quark matter is assumed from the outset---our data-driven Bayesian inference avoids such a presumption, even though the underlying NS EOS formalism remains the same.

The rest of the paper is organized as follows. In the next section, we shall outline our model framework and discuss the lower limit of hadron-quark transition density in cold neutron stars, considering indications from the Beam Energy Scan (BES) experiments at RHIC carried out by the STAR Collaboration recently. In section \ref{Results}, we present and discuss our results. Finally, a summary is given. Effects of the high-precision radius measurements on determining the crust-core transition density in neutron stars are given in the Appendix.

\section{A meta-model EOS for neutron star matter used in Bayesian analyses}\label{Theo}
For completeness and ease of our discussions, we recall in the following several main aspects of our approach. More detailed discussions on the physics justification, technical details and examples of applications can be found in our earlier publications (e.g.
\citep{zhang2018combined,Zhang:2019fog,Zhang:2021xdt,xie2019bayesian,xie2020bayesian,xie2021bayesian,Zhang:2023wqj,Zhang:2024npg,Xie:2024mxu,Li:2024imk,Xavier}) and reviews \citep{LiEPJA,Li:2021thg}. Compared to our earlier work within Bayesian statistical framework using the meta-model EOS with a first-order hadron-quark phase transition \citep{xie2021bayesian,Xavier}, as we shall discuss in detail the most important new physics and interesting results are (1) those related to using a varying precision of radius measurements in anticipation of coming observations and (2) effects of varying the prior range of the predicted hadron-quark transition density in comparison with findings of the recent BES/RHIC experiments. To avoid unnecessary repetitions, we shall make our summary here as brief as possible. 

\subsection{A meta-model EOS for neutron star matter}\label{meos}
In the CSS NS EOS model \citep{Alford:2013aca,zdunik2013}, the NS inner core of quark matter is connected to its outer core of hadronic matter through a first-order phase transition according to 
\begin{equation}
\varepsilon(p)= \begin{cases}\varepsilon_{\mathrm{HM}}(p) & \rho<\rho_{t} \\ \varepsilon_{\mathrm{HM}}\left(p_{t}\right)+\Delta \varepsilon+C_{\mathrm{qm}}^{-2}\left(p-p_{t}\right) & \rho>\rho_{t}\end{cases}
\end{equation}
where $\varepsilon_{\mathrm{HM}}(p)$ is the energy density of hadronic matter (HM) at pressure $p$, $p_t$ is the pressure at the transition density $\rho_t$, $\Delta \varepsilon$ describes the strength of the phase transition, and the speed of sound squared $C_{\rm{qm}}^2$ quantifies the stiffness of quark matter. Unless otherwise specified, the $C_{\rm{qm}}^2$ is measured in unit $c^2$ in the following discussions. These three parameters are generated randomly within their prior ranges specified in Table \ref{tab-prior}. Effectively, the CSS model is a template for generating quark matter EOS, i.e., a meta-model. It has been known that a CSS in QM well-approximated numerical calculations using the Nambu--Jona-Lasinio (NJL) model. Various other models have an approximately CSS in QM at densities relevant to NS, such as a bag model \cite{Zdunik:2000xx} and nonlocal NJL models \cite{nlNJL_cssContrera, nlNJL_cssShahrbaf}. Thus, the CSS model provides a simple approximation to analyze the effects of a first-order phase transition to QM. While signatures of the latter may have never been firmly detected, the simplicity and flexibility of the CSS model in mimicking various hadron-quark transition density and QM stiffness are sufficient for the stated purposes of this study. The results will be useful as a reference for future studies with more sophisticated hadron-quark transition models, e.g., with smooth crossovers. 

Similarly, we adopt a meta-model for the EOS of $npe\mu$ matter at $\beta-$equilibrium starting from parameterizing the binding energy per nucleon $E(\rho,\delta)$ in neutron-rich matter at nucleon density $\rho=\rho_n+\rho_p$ and isospin asymmetry $\delta\equiv (\rho_n-\rho_p)/\rho$ \citep{bombaci1991asymmetric}
\begin{equation}\label{eos}
E(\rho,\delta)=E_0(\rho)+E_{\rm{sym}}(\rho)\cdot \delta ^{2} +\mathcal{O}(\delta^4)
\end{equation}
where $E_0(\rho)$ is the symmetric nuclear matter (SNM) EOS and $E_{\rm{sym}}(\rho)$ is nuclear symmetry energy at density $\rho$. 

\begin{table}[htbp]
\centering
\caption{Prior ranges of the EOS parameters in units of MeV and the critical density $\rho_t/\rho_0$ for hadron-quark phase transition.}\label{tab-prior}
\begin{tabular}{lcc}
\hline\hline
Parameters & Lower limit & Upper limit \\
\hline
$K_0$ & 220 & 260 \\
$J_0$ & -400 & 400 \\
$K_{\mathrm{sym}}$ & -400 & 100 \\
$J_{\mathrm{sym}}$ & -200 & 800 \\
$L$ & 30 & 90 \\
$E_{\mathrm{sym}}(\rho_0)$ & 28.5 & 34.9 \\
$\Delta\epsilon/\epsilon_t$ & 0.2 & 1.0 \\
$C^2_{\rm qm}/c^2$ & 0.0 & 1.0 \\
$\rho_t/\rho_0$ (case A) & 1.0 & 6.0 \\
$\rho_t/\rho_0$ (case B) & 3.0 & 6.0 \\
\hline
\end{tabular}
\end{table}

The $E_0(\rho)$ and $E_{\rm{sym}}(\rho)$ can be parameterized as
\begin{eqnarray}\label{E0para}
  E_{0}(\rho)&=&E_0(\rho_0)+\frac{K_0}{2}(\frac{\rho-\rho_0}{3\rho_0})^2+\frac{J_0}{6}(\frac{\rho-\rho_0}{3\rho_0})^3,\\
  E_{\rm{sym}}(\rho)&=&E_{\rm{sym}}(\rho_0)+L(\frac{\rho-\rho_0}{3\rho_0})+\frac{K_{\rm{sym}}}{2}(\frac{\rho-\rho_0}{3\rho_0})^2
  +\frac{J_{\rm{sym}}}{6}(\frac{\rho-\rho_0}{3\rho_0})^3\label{Esympara},
\end{eqnarray}
where $E_0(\rho_0)=-16$ MeV at the SNM saturation density $\rho_0=0.16/\rm{fm}^3$. The coefficients $K_0$ and $J_0$ defined as
\begin{equation}
    K_0=9\rho_0^2[\partial^2 E_0(\rho)/\partial\rho^2]|_{\rho=\rho_0},\\\
    J_0=27\rho_0^3[\partial^3 E_0(\rho)/\partial\rho^3]|_{\rho=\rho_0} 
\end{equation}
are the SNM incompressibility and skewness, respectively. The $E_{\rm{sym}}(\rho_0)$, 
\begin{equation}
    L=3\rho_0[\partial E_{\rm{sym}}(\rho)/\partial\rho]|_{\rho=\rho_0},\\\ K_{\rm{sym}}=9\rho_0^2[\partial^2 E_{\rm{sym}}(\rho)/\partial\rho^2]|_{\rho=\rho_0},\\\
    J_{\rm{sym}}=27\rho_0^3[\partial^3 E_{\rm{sym}}(\rho)/\partial\rho^3]|_{\rho=\rho_0} 
\end{equation}
are the magnitude, slope, curvature, and skewness of nuclear symmetry energy at $\rho_0$, respectively. 
While these coefficients are all defined at $\rho_0$, the higher-order derivatives characterize the behaviors of  
$E_0(\rho)$ and $E_{\rm{sym}}(\rho)$ at densities significantly away from $\rho_0$ on both the super-saturation and sub-saturation sides, respectively. In particular, the skewness parameters $J_0$ and $J_{\rm{sym}}$ characterize the stiffness of 
$E_0(\rho)$ and $E_{\rm{sym}}(\rho)$ around $(3-4)\rho_0$ \citep{Xie:2020kta}, while $K_{\rm{sym}}$ and $L$ characterize the stiffness of $E_{\rm{sym}}(\rho)$ around $(1-3)\rho_0$ and $\rho_0$, respectively. They are important around the lower limit of the predicted hadron-quark transition density $\rho_t$. At sub-saturation densities, as we shall demonstrate, the higher-order parameters are also important for determining NS crust-core transition properties. 
As these high-order EOS parameters and the $\rho_t$ are already very uncertain with large prior ranges, it is unnecessary to introduce more parameters to describe the hadronic EOS at even higher densities. The prior uncertainty ranges of the nine EOS parameters listed in Table \ref{tab-prior} are mostly based on the knowledge the nuclear astrophysics community has accumulated over the last 40 years from analyzing terrestrial experiments, astrophysical observations, and extensive theoretical studies, see, e.g., Refs.\ \citep{LiEPJA,oertel2017equations,Esym-review,Garg18,Li:2018lpy} for recent reviews. Generally speaking, parameters characterizing nuclear symmetry energy around $\rho_0$ (e.g., $L$) are relatively well determined, while those characterizing its high-density behaviours are still poorly known. 

We note that the prior ranges adopted for some EOS parameters in Table~\ref{tab-prior} are intentionally broader than the most probable values suggested by certain existing experiments. These choices are made to be more inclusive and consistent with historically common ranges, with the expectation that the new data incorporated in the Bayesian analysis will independently constrain and reduce the associated uncertainties. This approach is motivated in part by the absence of a clear community consensus on essentially all of the EOS parameters listed. For example, based on approximately 100 independent analyses over the past two decades using diverse data sets, a fiducial value of $L \approx 60 \pm 20$~MeV has been extracted~\citep{Li:2013ola,oertel2017equations,Li:2025xio}. This value is in excellent agreement with the $\chi$EFT (chiral effective field theory) prediction of $L = 59.8 \pm 4.1$~MeV~\citep{drischler2020well}. Nevertheless, some analyses of several well-known experiments favor $L$ values that deviate significantly from this fiducial range. Moreover, predictions from more than 500 existing energy density functionals exhibit a broad dispersion around the fiducial value~\citep{Li:2025xio}. Although many of these functionals are disfavored or ruled out by various experimental constraints, they continue to be widely employed in the literature for different applications, albeit at the risk of confusion.

The pressure in $npe\mu$ matter at $\beta-$equilibrium can be calculated from
\begin{equation}\label{pressure}
  P(\rho, \delta) = \rho^2 \frac{\rm{d}\epsilon_{\rm{HM}}(\rho,\delta)/\rho}{\rm{d}\rho},
\end{equation}
where $\epsilon_{\rm{HM}}(\rho, \delta) = \rho [E(\rho,\delta)+M_N]+ \epsilon_l(\rho, \delta)$ is the energy density of NS matter with $\epsilon_l(\rho, \delta)$ the energy density of leptons. The latter is determined using the non-interacting Fermi gas model \citep{oppenheimer1939massive}. 
The density profile of isospin asymmetry $\delta(\rho)$ is obtained by using the $\beta$-equilibrium condition $\mu_n-\mu_p=\mu_e=\mu_\mu\approx4\delta E_{\rm{sym}}(\rho)$ and the charge neutrality requirement $\rho_p=\rho_e+\rho_\mu$. Here the chemical potential $\mu_i$ for a particle $i$ is calculated from the energy density via
$\mu_i=\partial\epsilon(\rho,\delta)/\partial\rho_i.$ Once the $\delta(\rho)$ is obtained, the pressure of $npe\mu$ matter becomes a function of density only (barotropic). 

The outer core EOS described above is then connected with the crust EOS at a crust-core transition density $\rho_{cc}$ consistently determined by examining when the outer core EOS becomes thermodynamically unstable 
against spinodal decomposition by forming clusters
\citep{Lattimer:2006xb,kubis2007nuclear,Xu:2009vi}. 
With the 6 nuclear matter EOS parameters randomly generated uniformly within their prior ranges, the resulting prior distribution of $\rho_{cc}$ is non-uniform around the fiducial value of $\rho_{cc}$=0.08 fm$^{-3}$ \citep{baym1971ground}. 
Our numerical analyses in the Appendix indicate that (1) with all other uncertainties fixed, high-precision radius measurements of canonical NSs are more useful for studying new physics associated with NS crust and constrain more precisely the $\rho_{cc}$, (2) high-precision radius measurements of massive NSs are more useful for investigating the nature and EOS of NS inner cores with little influence from existing uncertainties associated with NS crusts. This is mainly because even when the $\sigma$ becomes smaller than the thickness of NS crust ($\sim 1.0$ km) in measuring both $R_{1.4}$ and $R_{2.0}$, the crustal mass fraction is much smaller in the NS with mass 2.0 M$_{\odot}$. Thus, in the main text, we will focus on using high-precision $R_{2.0}$ data to probe properties of quark matter possibly existing in inner cores of NSs. 

We adopt the Negele-Vautherin (NV) EOS \citep{negele1973neutron} for the inner crust and the Baym-Pethick-Sutherland (BPS) EOS \citep{baym1971ground} for the outer crust. Finally, the complete NS EOS in the form of pressure versus energy density $P(\epsilon)$ is used in solving the Tolman-Oppenheimer-Volkoff (TOV) equations 
\citep{tolman1934effect,oppenheimer1939massive}.

In the context of Bayesian inference of NS data, virtually all EOS models can be considered meta-models, i.e., templates for constructing models \citep{Mar18,zhang2018combined,antic2019quantifying,Burrello:2025jay,Koehn:2024set,Klausner:2025ucq,
Klausner:2024jgu}. Each EOS model introduces parameters at different stages, either in microscopic/phenomenological constructions or in sampling large functional spaces via Gaussian processes. The number and ranges of these parameters define the model space, with varying complexity and potential limitations. A fundamental property of the TOV equations is their intrinsic composition degeneracy \citep{Cai:2025nxn}: any EOS specified as $P(\epsilon)$, regardless of its underlying constituents, produces the same mass-radius sequence. Thus, one can perform Bayesian inference directly on parametric or non-parametric $P(\epsilon)$ relations without knowing the composition. Posterior PDFs then provide constraints on the overall stiffness (e.g., the speed of sound) but do not directly reveal NS composition.  

In Appendix A of Ref. \citep{Cai:2024oom}, by comparing mass-radius sequences predicted by 284 so-called realistic NS EOSs covering various categories of theories in the literature with predictions using $10^5$ meta-model EOSs, we found that the meta-model EOSs can mimic most of the existing realistic NS EOSs. For the purposes of this work, this seemingly simple meta-model is as valuable as more elaborate EOS constructions used in the literature. A similar finding was made in Ref. \citep{Pro} using 40435 EOSs generated by another meta-model.

\subsection{Outlines of Bayesian analyses}\label{bayse}
As discussed in detail in our previous publications \citep{xie2019bayesian,xie2020bayesian,xie2021bayesian,Xie:2024mxu,Li:2024imk,Xavier}, we perform the standard Bayesian analyses according to Bayes's theorem
\begin{equation}\label{Bay1}
P({\cal M}|D) = \frac{P(D|{\cal M}) P({\cal M})}{\int P(D|{\cal M}) P({\cal M})d\cal M}.
\end{equation}
Here $P({\cal M}|D)$ represents the posterior probability of the model $\cal M$ (described here by the nine meta-model EOS parameters discussed above) given the dataset $D$. Meanwhile, $P(D|{\cal M})$ is the likelihood function that a given theoretical model $\cal M$ predicts the data $D$, and $P({\cal M})$ is the prior probability of the model $\cal M$ before comparing its prediction with the observational data. The denominator in Eq. (\ref{Bay1}) is a normalization constant. 
Effectively, our total likelihood function can be formally written as
\begin{equation}\label{Likelihood}
  P(D|{\cal M}) = P_{\rm{filter}} \times P_{\rm{mass,max}} \times P_\mathrm{R}
\end{equation}
where $P_{\rm{filter}}$ and $P_{\rm{mass,max}}$ indicate that the generated EOSs must satisfy the following conditions: (i) The crust-core transition pressure remains positive; (ii) The thermodynamic stability condition, $\rm{d}P/\rm{d}\epsilon\geq0$, holds at all densities; (iii) The causality condition is upheld at all densities; (iv) The generated NS EOS should be sufficiently stiff to support NSs at least as massive as 1.97 M$_{\odot}$ (i.e., the minimum M$_{\rm{TOV}}$ which is the maximum mass a given EOS can support). It is based on the mass observations by Antoniadis et al. of PSR J0348+0432 with $M = 2.01 \pm 0.04$ M$_\odot$ \cite{Antoniadis:2013pzd}. This condition was used in deriving constraints on the radius $R_{1.4}$ for canonical NSs from GW170817 by the LIGO/VIRGO Collaborations \cite{abbott2018gw170817}. As we use their result for $R_{1.4}$ as a basis for our study here, to be consistent, we thus adopt the 1.97 M$_{\odot}$ as the minimum M$_{\rm{TOV}}$.
Individual effects of varying the minimum M$_{\rm{TOV}}$ from 1.97 to 2.17 M$_{\odot}$ and its error bar as well as the way to implement them (sharp-cut off or as a Gaussian function), and turning on/off the causality condition on inferring the PDFs of EOS parameters for a given set of NS radius data were studied extensively by us earlier \citep{xie2019bayesian}. Indeed, enforcing differently the $P_{\rm{filter}}$ and $ P_{\rm{mass,max}}$ may lead to appreciable and non-trivial modifications to the posterior PDFs of some EOS parameters. Nevertheless, they have no effect on the conclusions of this work, as we will keep them the same while varying only the precision $\sigma_{\mathrm{obs},j}$ of the radius measurement in the radius likelihood function $P_\mathrm{R}$
\begin{equation}\label{Likelihood-R}
 P_\mathrm{R}[D(R_{1,2,\cdots N})|{\cal M}(p_{1,2,\cdots N})]
 =\prod_{j=1}^{N}\frac{1}{\sqrt{2\pi}\sigma_{\mathrm{obs},j}}\exp[-\frac{(R_{\mathrm{th},j}-R_{\mathrm{obs},j})^{2}}{2\sigma_{\mathrm{obs},j}^{2}}],
\end{equation}
where $R_{\mathrm{obs},j}$ is the observational result while $R_{\mathrm{th},j}$ is the theoretical prediction for the target $j$ ranging from 1 to the total observation number N. 

In our Bayesian analyses, the Metropolis-Hastings algorithm is utilized in the Markov Chain Monte Carlo (MCMC) process to generate posterior PDFs in the multi-parameter EOS space. Detailed and quantitative demonstrations and analyses of burn-in steps, convergence, and autocorrelation can be  
found in the Appendix of our previous work \citep{Li:2024imk} using essentially the same code. In particular, after 15,000 burn-in steps, we use 600,000 subsequent steps by each runner for calculating the posterior PDFs. Gradually, fewer steps are accepted in the MCMC process with higher precision $\sigma$. For example,  
the average acceptance rate in our MCMC process for $R_{1.4}=11.9$ km with $\sigma=1.0$ km and 0.1 km is about 28\% and 6\%, respectively. The results presented here use 24 runners for $\sigma=1.0$ km and 0.5 km, 
but 48 runners for higher precision. The relatively larger statistical fluctuations in some of our results with $\sigma=0.1$ km are due to the smaller acceptance rate. Nevertheless, this does not affect any of our qualitative conclusions.

\subsection{Limitations of the NS Meta-Model EOS and Necessary Cautions in Interpreting Bayesian Analyses}

It is important to recognize that the meta-model EOS introduced above has inherent limitations. Nevertheless, for the goals of this work, this EOS is scientifically sound, physically sufficient, and provides the flexibility needed to efficiently perform computationally expensive Bayesian analyses. At the same time, several cautions are necessary when interpreting the results. In particular, we emphasize the following points:

\begin{itemize}
    \item We assume that NS cores are composed of $npe\mu$ matter in $\beta-$equilibrium (outer core) and may also contain deconfined quarks in the inner core. A primary goal of this study is to evaluate the relative probability of quark matter cores, without assuming a priori that quarks are present. Within the Bayesian framework, we infer the probability density distributions of quark matter mass fractions and their radial extent. Other possible particles (e.g., heavy baryons, nucleon resonances such as $\Delta(1232)$ and $N^*(1440)$, hyperons, pions, kaons, or dark matter) are not included for three reasons: 

    (1) As with quark matter, there is currently no compelling observational evidence for these particles in NSs. For example, Ref.~\citep{Huang1} investigated whether current mass, radius, and tidal deformability measurements can constrain the presence of hyperons using two Bayesian models (with and without hyperons). Current observations cannot distinguish between these scenarios, though simulations of future high-precision data could provide evidence via Bayes factors. It was found that 
    even with 2\% uncertainties in mass and radius using mock data, only a limited number of measurements (six NSs) may be insufficient to  confirm the presence of hyperons. 

    (2) Effects of additional particles are partially absorbed in the large uncertainties (prior ranges) of the meta-model parameters for hadronic matter. For instance, our previous study \citep{Cai-D} on the ``$\Delta$-puzzle" shows that the critical densities at which the four charge states of $\Delta(1232)$ appear depend strongly on the slope $L$ of the nuclear symmetry energy, couplings of $\Delta$ with mesons (especially the isovector $\rho$ meson), and the in-medium masses of $\Delta$ resonances--all poorly constrained. Similar findings are reviewed in Ref.~\citep{Drago}. 

    (3) No published NS EOS study considers all possible particle species simultaneously. All analyses, including ours, have caveats. In this work, focusing on the relative probability of quark matter cores, it is scientifically and physically sound to adopt a ``minimal" model for hybrid stars, consisting solely of $npe\mu$ matter in $\beta-$equilibrium surrounding a possible quark matter inner core. This provides a well-constrained reference with minimal uncertainties, consistent with existing observations. 

    \item  The prior ranges of EOS parameters are informed by experimental and theoretical nuclear physics studies over the past four decades. High-density parameters remain poorly constrained, motivating future astrophysical and terrestrial measurements. Bayesian analyses using mock high-precision NS radius data aim to quantify how such future observations can reduce uncertainties in these parameters. We emphasize that Bayesian results are naturally prior-dependent. By incorporating physics insights, such as BES/RHIC experimental indications for the hadron-quark transition density in hot dense matter, the priors guide inference meaningfully. Purely statistical analysis with overly broad priors can yield mathematically most probable values that are physically unrealistic, particularly if data are insufficient to constrain parameter correlations. This is an intrinsic limitation of Bayesian inference, requiring careful interpretation. 

    \item Given the limited experimental constraints and model dependence of high-density EOS parameters, the prior ranges are chosen to encompass diverse predictions from microscopic many-body theories and phenomenological models. Wide priors do not dilute predictions but include them comprehensively. The posterior PDFs inferred from NS observational data provide the key microscopic information about NSs. As we demonstrate, high-precision NS radius measurements can significantly narrow the broad prior PDFs of high-density hadronic EOS parameters, yielding meaningful physical insights.
\end{itemize}

\subsection{Lower Limit of the Hadron-Quark Transition Density in Cold Neutron Stars}\label{BES}

Among the many open questions in neutron star physics, one of the most critical and least constrained is the hadron-quark transition density $\rho_t$ associated with a possible first-order phase transition in dense matter. Depending on the underlying theoretical framework, predictions for $\rho_t$ in cold $\beta$-equilibrated matter span a broad range, e.g., $(1.5$--$3.0)\rho_0$ from percolation arguments, $(2.0$--$4.0)\rho_0$ within the Nambu-Jona-Lasinio model, and $(2.0$--$5.0)\rho_0$ in quarkyonic matter scenarios \citep{Alford:2004pf,Baym:2017whm,Annala:2019puf,Montana:2018bkb,Blaschke:2013ana,Fischer:2017lag}. Interestingly, most of the recent Bayesian analyses of NS observational data since GW170817 tend to favor the lower end of these intervals, typically adopting $\rho_t^{\rm low}=1.0\rho_0$ as the lower bound of the prior in Bayesian inference, see, e.g. refs.  \citep{Li:2021crp,Miao:2020yjk,Christian:2023hez,Christian:2025dhe,Xia:2024wpz,Naseri:2024rby,Albino:2024ymc,Somasundaram_2022,Albino:2025puc,Ayriyan:2025rub}. The hope is that future precision NS data will further narrow down the allowed range of $\rho_t$ or determine its most probable value with quantified uncertainties.

In parallel, the relativistic heavy-ion collision program has been actively exploring hot, dense QCD matter and the emergence or disappearance of quark-gluon plasma (QGP) signatures \citep{Fukushima:2010bq,Busza:2018rrf,Xu:2022mqn,Chen24}. In particular, the BES program at RHIC by the STAR Collaboration \citep{STAR1,STAR3,STAR2} reports strong indications that QGP signatures disappear in Au+Au collisions at $\sqrt{s_{NN}} = 3$~GeV. At this energy, collective flow observables are well reproduced by hadronic transport models with nuclear mean-field potentials, suggesting that the dense medium formed is dominantly hadronic \citep{STAR1,STAR3}. In addition, proton cumulant measurements indicate that if a QCD critical region is formed, it must lie at energies $\sqrt{s_{NN}} > 3$~GeV. At $\sqrt{s_{NN}} = 4.5$~GeV, clear signatures of partonic collectivity re-emerge \citep{STAR:2025owm}. Within an updated relativistic transport model (ART) \citep{ART}, the maximum central baryon density reached in $\sqrt{s_{NN}} = 3$~GeV Au+Au collisions is $\sim 3.6\rho_0$ \citep{Yong22}. Together, these findings provide strong circumstantial evidence that the hadron-quark transition density in hot dense matter should exceed $\rho_t^{\rm low}\approx 3.6\rho_0$.

A natural question arises: To what extent is the matter produced in heavy-ion collisions comparable to that in NS cores, given the differences in temperature, isospin asymmetry, equilibrium conditions, and composition? While the QGP observed in heavy-ion collisions is not realized in cold neutron stars, both systems occupy distinct regions of the same QCD phase diagram. They are connected through the continuous first-order hadron-quark transition boundary in the $T$--$\mu_B$ (temperature-baryon chemical potential) or $T$--$\rho$ plane \citep{Baym:2017whm,Jako,Blacker:2023afl}. This boundary is typically arc-shaped, rising in density as temperature decreases starting from the critical end point (CEP), as illustrated in Refs.~\citep{Alford:2007xm,Kumar:2024owe}. Calculations within a chiral mean-field model \citep{Kent} show that the transition lines for hot isospin-symmetric matter and for matter in $\beta$-equilibrium relevant to NS mergers nearly coincide down to $T\approx100$~MeV. Below that, and toward $T=0$, the transition density in neutron stars is only slightly lower than in isospin-symmetric matter. Thus, within such models, it is generally unlikely for $\rho_t$ in cold neutron stars to be \emph{lower} than that in the hot matter created at BES/RHIC energies. 

That said, this trend is model-dependent and can potentially be altered by isospin effects, trapped neutrinos, Coulomb and surface energies in mixed phases, or choices in the EOS modeling of quark and hadronic matter. In particular, the isospin dependence of the transition boundary and the CEP location on the QCD phase diagram remains an open problem with active ongoing work \citep{Kent,Sag09,David10,Yao:2023yda}.

Motivated by these theoretical and experimental developments, we explore how our Bayesian inference results depend on the choice of $\rho_t^{\rm low}$ in the prior. Importantly, our approach does \emph{not} assume a one-to-one equivalence between the matter created in heavy-ion collisions and that in neutron stars, nor does it rely on the dynamical history of either system. Rather, it is based on the underlying continuity of QCD matter under changing thermodynamic conditions, of which cold neutron star matter is simply the zero-temperature limit.

While it is compelling to examine how $\rho_t^{\rm low}$ impacts astrophysical inference, we stress that extracting $\rho_t$ from BES/STAR data remains nontrivial \citep{STAR1,STAR3,STAR2,Yong22}. To remain conservative, we therefore adopt $(3$--$6)\rho_0$ as the BES/RHIC-informed prior range for $\rho_t$ in cold neutron stars. Determining the transition boundary and the CEP on the QCD phase diagram with high accuracy is a major objective of the current nuclear physics program \citep{Fukushima:2010bq,Busza:2018rrf,Xu:2022mqn,Chen24,Lovato:2022vgq,Sorensen:2023zkk,Almaalol:2022xwv}. High-precision neutron star radius measurements would provide a complementary and potentially decisive constraint on $\rho_t$ in the cold, dense regime, effectively extending the QCD phase diagram into the isospin dimension. In this work, we therefore compare Bayesian inference results obtained with two prior choices for $\rho_t$: $(1$--$6)\rho_0$ (case-A) and $(3$--$6)\rho_0$ (case-B), as summarized in Table~\ref{tab-prior}.

\section{Results and Discussions}\label{Results}
\subsection{Hybrid star EOS parameters inferred assuming a prior range $(1-6)\rho_0$ for the hadron-quark transition density $\rho_t$}
\begin{figure}[ht]
\centering
 \resizebox{1.\textwidth}{!}{
  \includegraphics{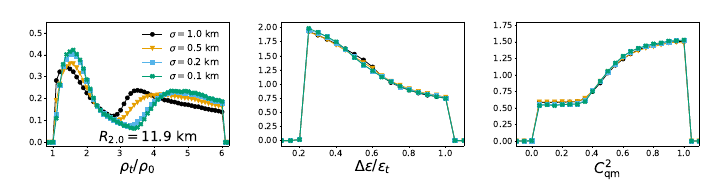}
  }
\setlength{\abovecaptionskip}{-0.5cm}
  \caption{Posterior PDFs of the three quark matter EOS parameters inferred from mock NS radius data of $R_{2.0}=11.9$ km, with the precision $\sigma$ varying from 1.0 to 0.1 km.}\label{PDF-QM-rt1}
\end{figure}

\begin{figure}[ht]
\centering
 \resizebox{1.\textwidth}{!}{
  \includegraphics{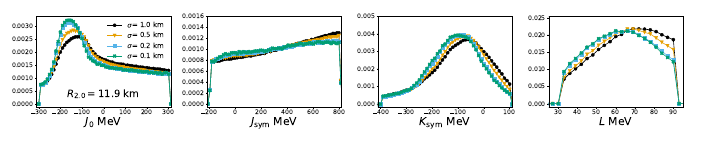}
  }
\setlength{\abovecaptionskip}{-0.7cm}
  \caption{(color online) Posterior PDFs of four hadronic EOS parameters inferred from the same data sets as in Fig. \ref{PDF-QM-rt1}.}\label{PDF-HM-rt1}
\end{figure}

In this subsection, we present results from our Bayesian analyses adopting the widely used fiducial prior range $(1-6)\rho_0$ (case A) for the hadron-quark transition density $\rho_t$ and the mock radius data $R_{2.0}=11.9$ km with $\sigma_{\mathrm{obs},j}$ varying from 1.0 to 0.1 km, respectively. Shown in Fig. \ref{PDF-QM-rt1} are the posterior PDFs of the three quark matter EOS parameters, while the PDFs of the four high-density hadronic EOS parameters are shown in Fig. \ref{PDF-HM-rt1}. The PDFs of two saturation-density parameters $K_0$ and $E_{\rm sym}(\rho_0)$ are not shown as they are much less important and interesting for this work. Among the interesting features observed, we emphasize the following 

\begin{itemize}
    \item The PDF($\rho_t$) generally has a major peak around $(1.7-2.0)\rho_0$ and a minor peak or a broad bump around $(3.0-5.0)\rho_0$. Their sharpness or width varies significantly as the precision $\sigma$ changes from 1.0 to 0.2 km, then stays more or less the same when $\sigma$ is further reduced. 
    
    \item Narrow widths of the mixed phase $\Delta\epsilon/\epsilon_t$ and very stiff quark matter EOSs with high $C^2_{\rm{qm}}$ values are preferred regardless of the precision $\sigma$. In particular, the approximately equally high probability for $C^2_{\rm{qm}}$ to be in the range of $(0.6-1.0)c^2$ corresponds to the major peak of PDF($\rho_t$) around $(1.7-2.0)\rho_0$ and the narrow width $\Delta\epsilon/\epsilon_t$ of the transition. It simply indicates that when the hadronic matter starts transitioning to quark matter at a rather low density, the EOS of quark matter is required to be very stiff to support NSs at least as massive as $1.97$ M$_{\odot}$ and the specified radius data. However, the precision $\Delta R$ has little effect on the PDFs of $\Delta\epsilon/\epsilon_t$ and $C^2_{\rm{qm}}$. 
    
    \item Generally speaking, the PDFs of the four most important parameters describing hadronic EOS have a small dependence on the precision $\sigma$. These results are consistent with our earlier findings in Ref. \citep{Li:2024imk} using the minimum NS EOS model without considering the hadron-quark phase transition. 
    \end{itemize}

\begin{table*}[htbp]
\centering
\caption{Mean values of nine EOS parameters inferred from the mock data $R_{2.0}=11.9$ km}
\label{r20}
\begin{tabular}{lcccc}
\toprule
\hline
Pars & $\sigma = 0.1$ & $\sigma = 0.2$ & $\sigma = 0.5$ & $\sigma = 1.0$ \\
\midrule
\hline
$J_0$ (MeV) & -33.9$\pm$161 & -33.0$\pm$160. & -24.4$\pm$158 & -13.7$\pm$159 \\
$K_0$ (MeV) & 240.$\pm$11.5 & 240.$\pm$11.5 & 240.$\pm$11.5 & 240.$\pm$11.5 \\
$J_{\rm{sym}}$ (MeV) & 327$\pm$285 & 329$\pm$286 & 339$\pm$287 & 347$\pm$287 \\
$K_{\rm{sym}}$ (MeV) & -121$\pm$103 & -118$\pm$103 & -106$\pm$106 & -100.$\pm$111. \\
$L$ (MeV) & 60.8$\pm$15.7 & 61.1$\pm$15.7 & 62.8$\pm$15.6 & 64.0$\pm$15.7 \\
$E_{\rm{sym}}(\rho_0)$ (MeV) & 31.8$\pm$1.85 & 31.8$\pm$1.85 & 31.7$\pm$1.85 & 31.7$\pm$1.85 \\
$\rho_t/\rho_0$ & 3.33$\pm$1.59 & 3.34$\pm$1.58 & 3.33$\pm$1.52 & 3.23$\pm$1.47 \\
$\Delta\epsilon/\epsilon_t$ & 0.525$\pm$0.224 & 0.526$\pm$0.224 & 0.530$\pm$0.224 & 0.526$\pm$0.223 \\
$C^2_{\rm{qm}}$ & 0.609$\pm$0.266 & 0.607$\pm$0.267 & 0.600$\pm$0.270 & 0.602$\pm$0.270 \\
\bottomrule
\hline
\end{tabular}
\end{table*}

To be more quantitative, shown in Table \ref{r20} are the mean values and standard deviations of all nine EOS parameters inferred. Since some of the PDFs have dual peaks or are flat, and the relevant Maximum {\it a Posterioris} (MaPs) can be seen obviously, while their confidence boundaries are not very useful for our purposes here, the MaPs and their confidence boundaries are not listed. Comparing the results from using $\sigma=1.0$ km and 0.1 km, we notice that the mean value of $J_0$ changes from $-13.7\pm 159$ MeV to $-33.9\pm 161$ MeV, the changes in $J_{\rm sym}$ and $K_{\rm sym}$ are small, while all others have basically no change within the statistical errors listed. Overall, for hybrid stars with a transition density to quark matter as low as $\rho^{\rm low}_t=\rho_0$, even a very precise measurement of $R_{2.0}$ does not provide particularly more stringent constraints on the nucleonic EOS parameters. Nevertheless, as we shall show in the next subsection, using $\rho^{\rm low}_t=3.0\rho_0$, the precision impact becomes more pronounced.

\begin{figure}[ht]
\centering
 \resizebox{0.8\textwidth}{!}{
  \includegraphics[trim={5mm, 5mm, 5mm, 28mm}, clip, width=16cm]{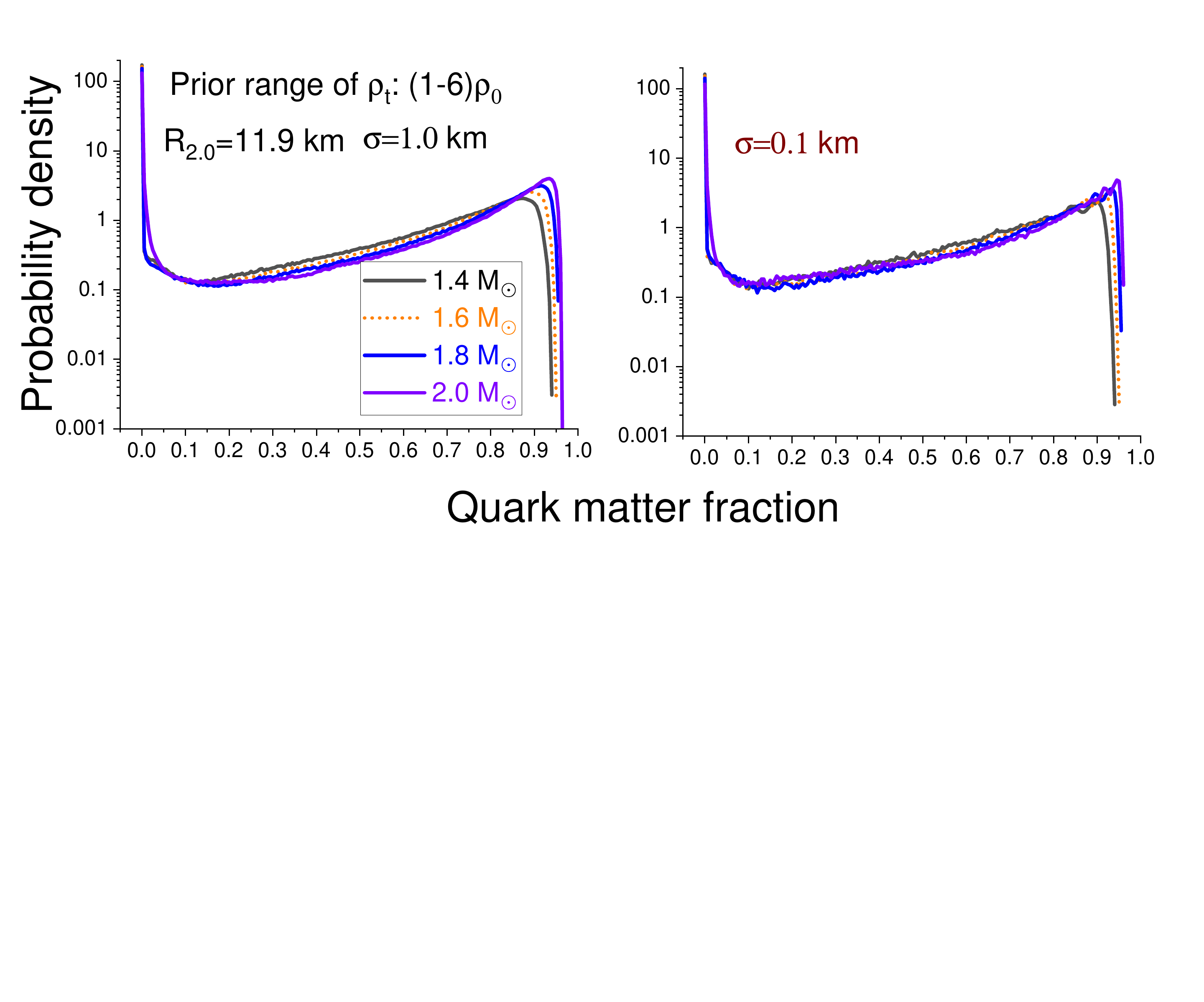}
  }
\setlength{\abovecaptionskip}{-5.3cm}
  \caption{color online) Quark matter mass fraction of hybrid stars inferred from the mock radius data of a massive NS of mass 2.0M$_{\odot}$ with a mean radius $R_{2.0}=11.9$ km and the precisions indicated.}\label{QMR20}
\end{figure}
To learn more about quark matter cores, we now examine in Fig. \ref{QMR20} the probability density of the mass fraction $f_{\rm{QM}}$ of quark cores in a hybrid star over its total mass with $\sigma$ taken at 1.0 km (left) and 0.1 km (right), respectively. The quark matter mass is obtained by integrating the energy density from the center to the starting energy density $\epsilon_c=\epsilon_{\rm{HM}}(\rho_t) +\Delta \epsilon$ of the quark core. The following features are most interesting to notice: In both cases, the probability density has two peaks. The dominating peak at $f_{\rm{QM}}=0$ represents purely hadronic stars. Hybrid stars reach maximum $f_{\rm{QM}}$ values around 0.95, depending on their total mass. Obviously, the more massive ones have a higher chance of having more quark matter. The probability of the second peak around $f_{\rm{QM}}=0.9$ is generally about (20-80) times smaller than that of the first peak. The latter corresponds to the major peak of the PDF($\rho_t$) around $\rho_t=(1.7-2.0)\rho_0$ shown in Fig. \ref{PDF-QM-rt1}, while the broad shoulders between the two peaks are due to the wide PDF($\rho_t$) of the transition density $\rho_t$. Effects of precision $\sigma$ on inferring $f_{\rm{QM}}$ are very small in this case of using the prior range $(1-6)\rho_0$ for $\rho_t$. 

\subsection{Effects of the prior range of hadron-quark transition density $\rho_t$ on inferring hybrid star EOS parameters}
The results presented above from using the fiducial prior range $(1-6)\rho_0$ for $\rho_t$ (Case A) are certainly interesting, especially in demonstrating how the high-precision radius measurements may provide deeper insights into properties of hybrid stars. As we discussed earlier in Section \ref{BES}, it would also be interesting to investigate how the prior range of $\rho_t$, especially its lower limit $\rho^{\rm{low}}_t$, may affect what one can learn about hybrid stars from future high-precision NS radius data. Considering indications about $\rho^{\rm{low}}_t$ from relativistic heavy-ion collisions, in the following we present results obtained by using the prior range $(3.0-6.0)\rho_0$ for $\rho_t$ (Case B).

\begin{figure}[ht]
\centering
 \resizebox{1.\textwidth}{!}{
  \includegraphics[trim={5mm, 90mm, 5mm, 7mm}, clip, width=16cm]{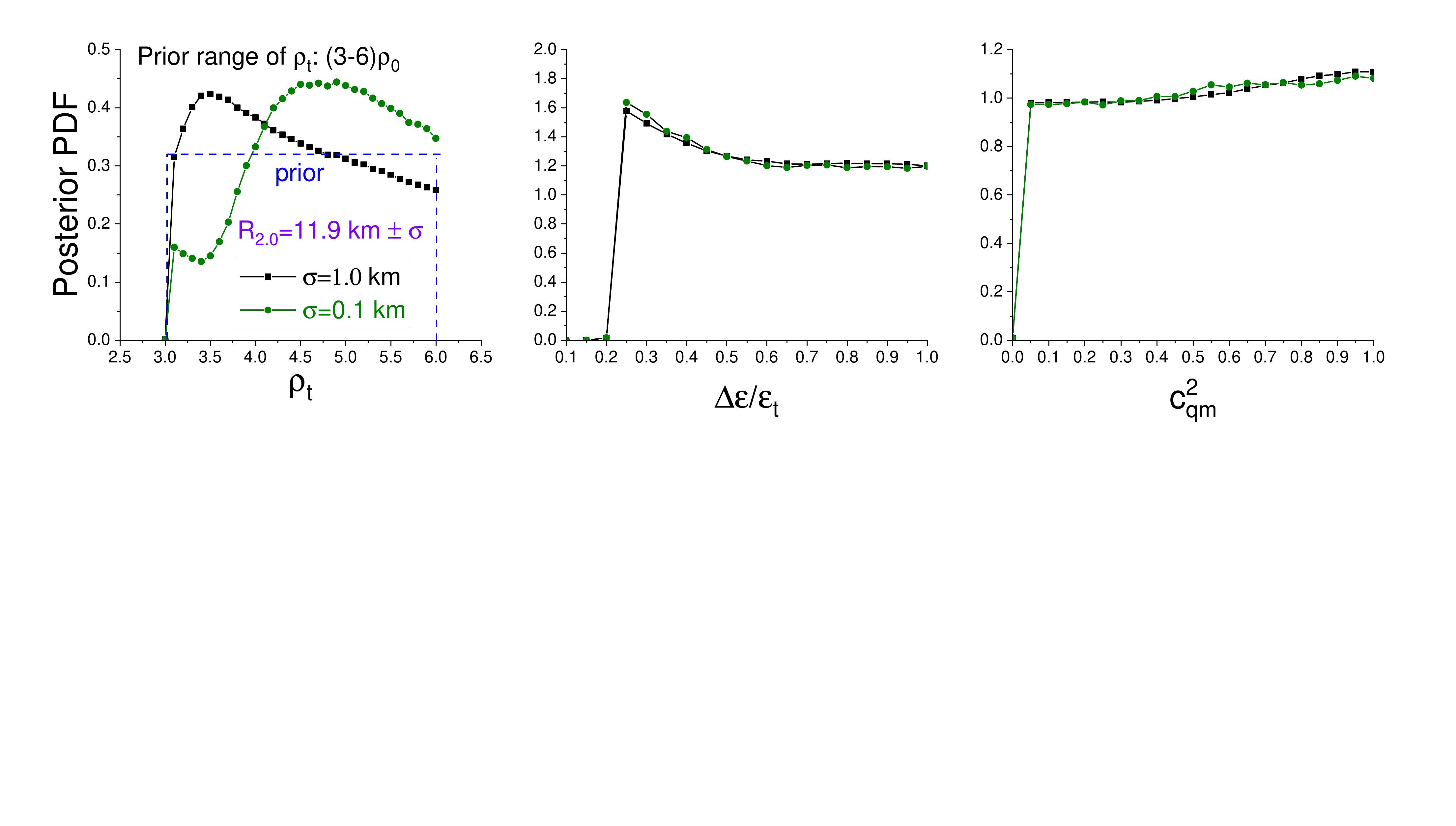}
  }
\setlength{\abovecaptionskip}{-3.4cm}
  \caption{(color online) Posterior PDFs of the three quark matter EOS parameters inferred from $R_{2.0}=11.9$ km data using $(3.0-6.0)\rho_0$ as the prior range for $\rho_t$ and precision $\sigma=1.0$ and $0.1$ km, respectively.}\label{QM36R20}
\end{figure}

Shown in Fig. \ref{QM36R20} are the posterior PDFs of the three quark matter EOS parameters inferred from $R_{2.0}=11.9$ km with a precision of $\sigma=1.0$ and $0.1$ km, respectively. Firstly, it is interesting to see that with $\sigma=1.0$ km, the MaP of $\rho_t$ is about $3.5\rho_0$, consistent with the $\rho^{\rm{low}}_t$ from analyzing the BES/STAR experiments. Moreover, as the precision improves to $\sigma=0.1$ km, the MaP of $\rho_t$ increases to about $4.7\rho_0$.
Clearly, comparing the results in Case A and Case B, the PDF and the MaP of $\rho_t$ are now much more sensitive to $\sigma$. Intuitively, since the average density $\rho_a$ of an NS scales with $M/R^3$, a small variation in its radius can lead to a big change in its $\rho_a$ and density profile. It is also well known that the correspondence between the NS radius and the underlying EOS, especially in the high-density region, is highly nonlinear. This is mainly because of the highly nonlinear nature of the TOV equations \citep{Cai:2023pkt,Cai:2025nxn}. The strong sensitivity of PDF($\rho_t$) to the precision $\sigma$ of NS radius measurements is thus understandable. We notice that the MaP values of $\rho_t$ in Case B are basically positions of the second peaks/bumps in the posterior PDFs($\rho_t$) obtained earlier in Case A (as shown in Fig. \ref{PDF-QM-rt1}). It indicates that the first peak in the PDF($\rho_t$) around $(1.7-2.0)\rho_0$ found earlier in Case A is statistically sufficient but not physically necessary to describe all NS observational data, if one believes in the indication from the BES/RHIC experiments. Fundamentally, pure statistics can not replace physics insights. Fortunately, Bayes' theorem naturally incorporates the latter partially through the priors used.

\begin{figure}[ht]
\hspace{-0.8cm}
\centering
 \resizebox{1.05\textwidth}{!}{
   \includegraphics{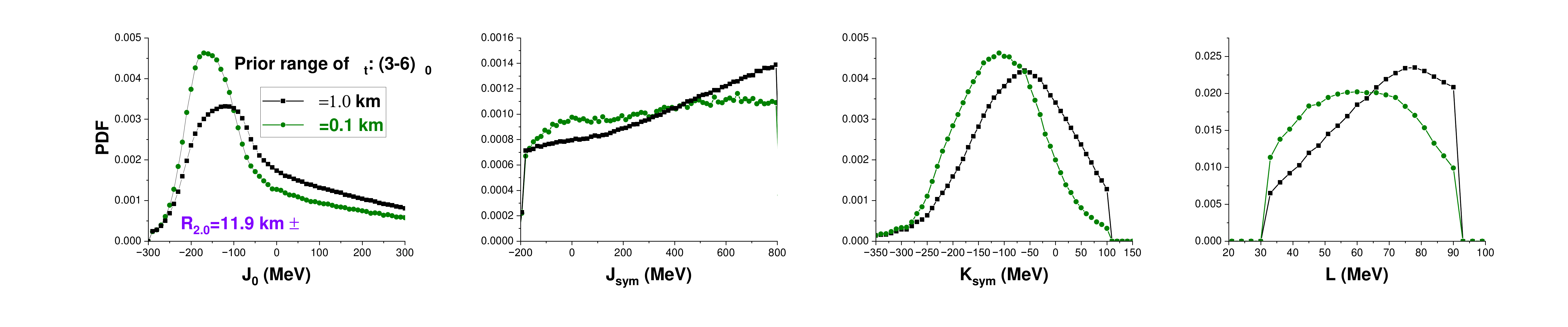}
  }
\setlength{\abovecaptionskip}{-1.cm}
  \caption{(color online) Same as in Fig. \ref{QM36R20} but for the posterior PDFs of the four hadronic matter EOS parameters.}\label{NM36R20}
\end{figure}

It is also very interesting to see that the precision $\sigma$ has essentially no effect on the posterior PDFs of quark matter properties quantified by $\Delta\epsilon$ and $C^2_{\rm{qm}}$ regardless of the prior range used for $\rho_t$ (in both Fig. \ref{QM36R20} and Fig. \ref{PDF-QM-rt1}). Moreover, the PDF($C^2_{\rm{qm}}$) is rather flat in its whole range, indicating that the NS radius data, regardless of its precision, does not constrain the quark matter stiffness. In our opinion, this is physical, although it may sound very disappointing to some people. In fact, it has been well known that the radii of canonical NSs are determined by the pressure at densities around $2\rho_0$ \citep{Lattimer:2000nx}. For massive NSs, the relevant density is expected to be higher \citep{Cai:2023gol}. More quantitatively, using a Skyrme Hartree-Fock energy density functional in Ref.~\citep{Fattoyev:2014pja}, $R_{1.4}$ and $R_{1.8}$ were found to correlate most strongly with nuclear symmetry energy around $1.5\rho_0$ and (2-3)$\rho_0$, respectively.
Therefore, with the most probable hadron-quark transition density $\rho_t$ as high as $3.5\rho_0$ with $\sigma=1.0$ km and $4.7\rho_0$ with $\sigma=0.1$ km in NSs with mass 2.0 M$_{\odot}$, the stiffness of hadronic (quark) matter is (NOT) expected to affect significantly the radii even for massive NSs. 

\begin{figure*}[ht]
\begin{center}
 \resizebox{0.9\textwidth}{!}{
  \includegraphics[trim={5mm, 370mm, 5mm, 34mm}, clip, width=16cm]{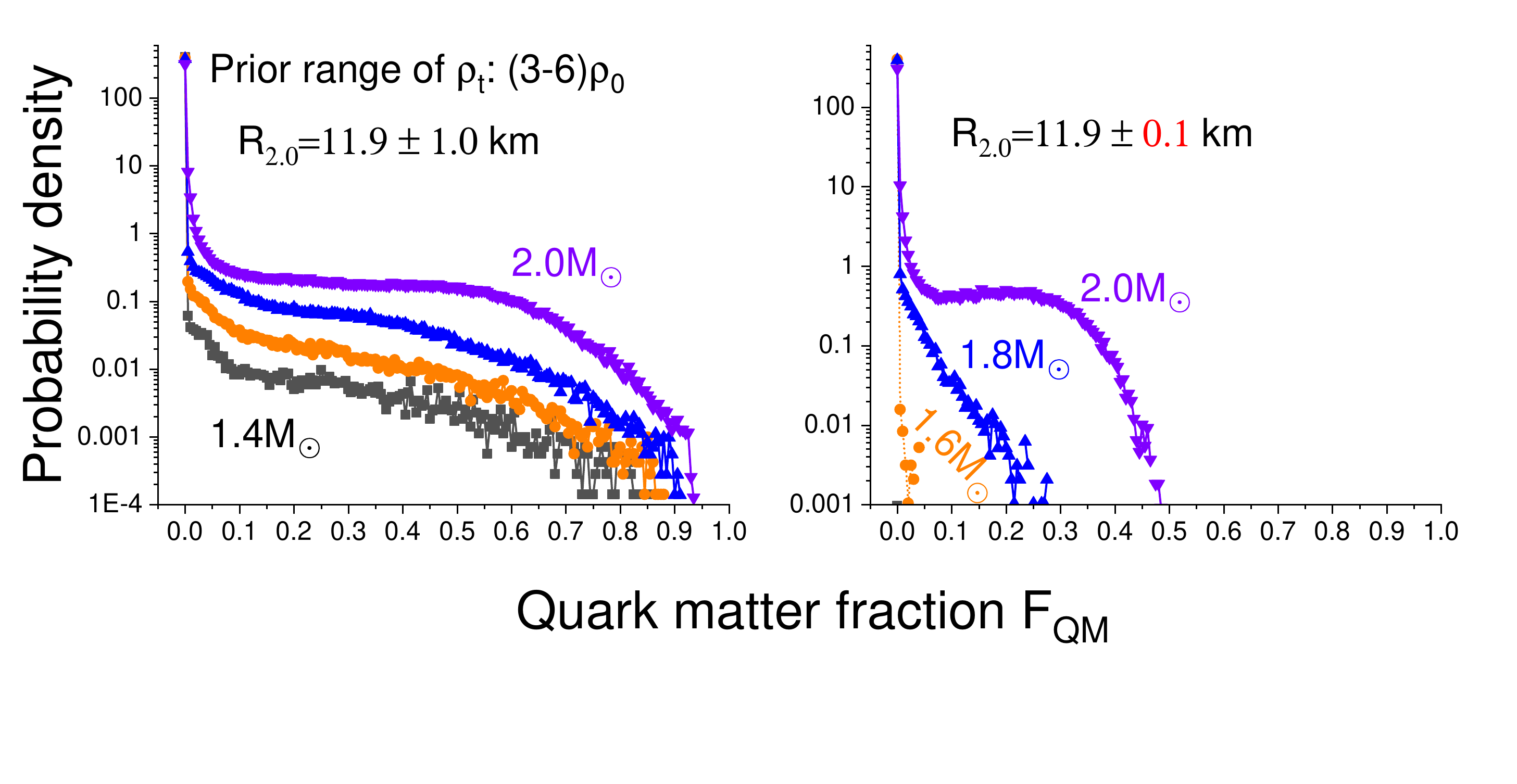}
  }
\setlength{\abovecaptionskip}{-1.6cm}
  \caption{(color online) Probability density of quark matter mass fraction inferred from $R_{2.0}=11.9$ km with $\sigma=1.0$ km (left) and 0.1 km (right), respectively.
}\label{fQM36}
\end{center}
\end{figure*}

\begin{figure*}[ht]
\begin{center}
 \resizebox{1.4\textwidth}{!}{
  \includegraphics[trim={5mm, 170mm, 5mm, 14mm}, clip, width=16cm]{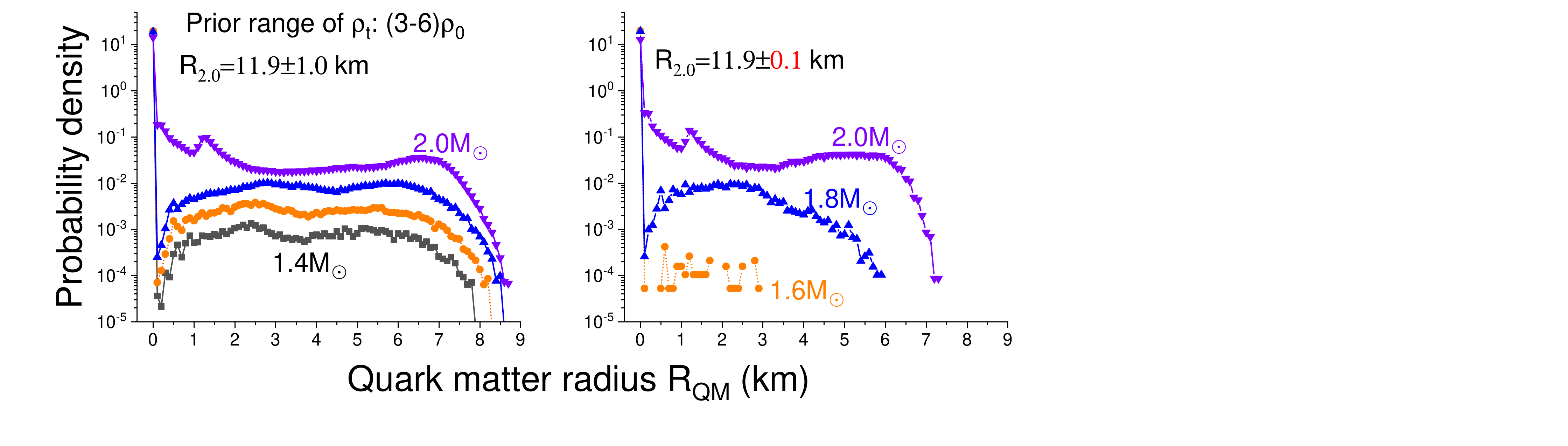}
  }
\setlength{\abovecaptionskip}{-1.cm}
  \caption{(color online) Distribution of quark matter radius inferred from $R_{2.0}=11.9$ km with $\sigma=1.0$ km (left) and 0.1 km (right), respectively.
   }\label{rQM36}
\end{center}
\end{figure*}

The above findings and discussions provide some hints about important roles of hadronic EOS parameters in determining NS radii. Then, one relevant question is how future high-precision NS radius measurements can improve our knowledge about high-density hadronic matter close to its interface with quark matter. To answer this question, shown in Fig. \ref{NM36R20} are the posterior PDFs of the four hadronic EOS parameters in Case B. Compared to the results shown earlier in Fig. \ref{PDF-HM-rt1} for Case A, the effects of $\sigma$ on all hadronic EOS parameters are qualitatively consistent. 
Quantitatively, a higher $\rho^{\rm low}_t$ leads to a greater sensitivity to the nucleonic part of NS EOS, albeit limited. It indicates again that NS radii are robust probes of the dense hadronic matter EOS around $(2.0-3.0)\rho_0$, not much affected by the uncertainty associated with the hadron-quark transition. Not surprisingly, the PDFs of $L$, $K_{\rm{sym}}$, and $J_0$ are appreciably more constrained with $\sigma=0.1$ km than with 1.0 km in both cases. 
However, the PDF of $J_{\rm{sym}}$ characterizing the $E_{\rm{sym}}(\rho)$ above about $3.0\rho_0$ remains unconstrained in its very large prior range. 

One interesting indication of Fig. \ref{QM36R20} is that the posterior PDF($\rho_t$) itself is sensitive to the precision $\sigma$ of radius measurements. As $\rho_t$ marks the end of the hadronic phase, we thus expect the $\sigma$ to affect significantly what one can infer about the quark matter fraction in hybrid stars as a result of mass conservation. To test this expectation, shown in Fig. \ref{fQM36} and Fig. \ref{rQM36} are the probability density of quark matter mass fraction and its size (measured by the radius $R_{\rm{QM}}$ where $\epsilon\geq\epsilon_{\rm{HM}}(\rho_t) +\Delta \epsilon$) inferred from $R_{2.0}=11.9$ km with $\sigma=1.0$ km (left) and 0.1 km (right), respectively. Most obviously and interestingly, because the strong sensitivity of PDF$(\rho_t)$ to the precision $\sigma$, the quark mass fractions obtained with 
$\sigma=0.1$ km and 1.0 km are very different. In particular, with $\sigma=0.1$ km because the most probable $\rho_t$ is as high as $4.7\rho_0$, only massive stars heavier than about 1.8 M$_{\odot}$ have an appreciable quark matter core. Nevertheless, we should note that the total probability of having a quark core is very small in any case studied here. More quantitatively, the probability of having (10-30)\% mass in the quark core of a 2.0 M$_{\odot}$ hybrid star is roughly three orders of magnitude smaller than that of having a purely hadronic star. It is seen from Fig. \ref{rQM36} that most of the quark matter are distributed around the center while some small amounts of them are spread out to as far as about 8.5 km away because the transition density $\rho_t$ has a broad distribution.

\section{Summary and Conclusions}

In this work, using mock high-precision NS radius data expected to become available in the near future from upcoming X-ray and gravitational-wave observatories, we inferred the posterior PDFs of meta-model EOS parameters for hybrid stars. The main conclusions of our analysis are summarized as follows:
\begin{itemize}
    \item High-precision NS radius measurements for massive NSs, provide strong constraints on the hadron-quark transition density $\rho_t$.
    
    \item The relatively low transition densities $\rho_t \approx (1.7$--$2.0)\rho_0$, favored in several previous analyses of existing NS observations, are incompatible with the recent BES/STAR experimental indications at RHIC. This tension suggests that such low values of $\rho_t$ may simply represent an effective combination of EOS parameters sufficient to explain current data, rather than a physically necessary feature.
    
    \item Adopting a prior range of $\rho_t \in (3.0$--$6.0)\rho_0$, consistent with BES/STAR constraints on hot dense matter, we find that high-precision $R_{2.0}$ measurements can significantly tighten constraints on the quark matter mass fraction and its radial extent, as well as several supradense hadronic EOS parameters, compared to current knowledge.
    
    \item While high-precision $R_{2.0}$ data are highly effective in constraining $\rho_t$, they are much less sensitive to the stiffness of quark matter, quantified by the squared speed of sound. This is fundamentally because $R_{2.0}$ is determined primarily by the pressure of hadronic matter at densities just below $\rho_t$. Since $\rho_t$ sets the maximum density of the hadronic phase, constraints on $R_{2.0}$ indirectly limit the quark matter mass fraction and core size through baryon number conservation.
    
    \item The strong dependence of what we can learn about superdense matter in hybrid stars on the precision of radius measurements should not be viewed as a limitation of the modeling. Rather, it reflects the intrinsically nonlinear structure of the TOV equations and the highly nontrivial mapping between the EOS and the mass-radius relation. As in many scientific disciplines, the physical insight gained is fundamentally limited by the precision of the observational tools available.
\end{itemize}

\noindent{\bf Acknowledgement:} We thank Nu Xu for a critical reading of the manuscript and his helpful comments. BAL and XG were supported in part by the U.S. Department of Energy, Office of Science, under Award Number DE-SC0013702, the CUSTIPEN (China-U.S. Theory Institute for Physics with Exotic Nuclei) under the US Department of Energy Grant No. DE-SC0009971. WJX was supported in part by the Shanxi Provincial Foundation for Returned Overseas Scholars under Grant No 20220037, the Natural Science Foundation of Shanxi Province under Grant No 20210302123085, the Open Project of Guangxi Key Laboratory of Nuclear Physics and Nuclear Technology, No. NLK2023-03 and the Central Government Guidance Funds for Local Scientific and Technological Development, China (No. Guike ZY22096024). NBZ is supported in part by the National Natural Science Foundation of China under Grant No. 12375120, the Zhishan Young Scholar of Southeast University under Grant No. 2242024RCB0013.

\appendix 
\vspace{-1cm}
\indent
\renewcommand\theequation{a\arabic{equation}}
\section*{\small A. Potential effects of high-precision NS radius data on crust-core transition density}\label{A1}
While the focus of this work is on the nature and EOS of supradense matter in the NS inner cores based on Bayesian analyses of expected high-precision NS radius data, it is also important to examine possible uncertainties in our work due to our poor knowledge about the crust-core transition at low densities. Moreover, it is interesting to know if high-precision NS radius data may help reduce these uncertainties. The thickness of NS crust is widely believed to be only about (10-15)\% of its radius (comparable with the observational uncertainty of existing NS radius measurements), and its mass fraction is less than 1.5\% for a canonical NS \cite{Xu:2009vi}. Nevertheless, many interesting physical processes and phenomena are predicted to occur inside and around the crust, see, e.g., Refs. \cite{Newton:2014iha,Sotani:2024mlb,Shchechilin:2024kjv,Davis:2024nda}, albeit with little observational evidence or constraint so far for most of them. It is known that employing different crust models can introduce an uncertainty of about 0.1 to 0.7 km (Table I of Ref. \cite{Fortin:2016hny}) to the radii of canonical NSs. It is relatively very large compared to the thickness of their crusts. 

\renewcommand*\figurename{\small a.Fig.}
\setcounter{figure}{0}
\begin{figure*}[ht]
\centering
 \resizebox{1.8\textwidth}{!}{
  \includegraphics[]{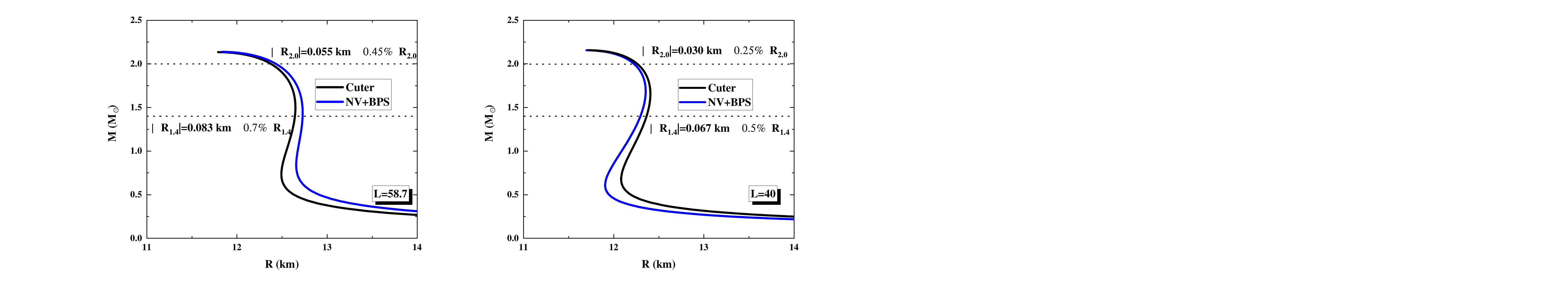}
  }
\setlength{\abovecaptionskip}{-0.7cm}   
  \caption{(color online) Comparisons of the mass-radius sequences using the same core EOS without a hadron-quark phase transition but different crustal EOSs: (1) the same NV+BPS (blue) and (2) the CUTER crustal EOS (black). Both panels use the EOS parameters $E_{\rm sym}(\rho_0)=31,7$ MeV, $K_{\rm sym}=-100$ MeV, $J_{\rm sym}=800$ MeV, $K_0=240$ MeV, and $J_0=-190$ MeV. The left (right) panel uses $L=58.7$ (40) MeV.}\label{cuter}
\end{figure*}

To the best of our knowledge, no existing model in the literature incorporates all proposed aspects of crustal physics, and the NV+BPS crustal EOS adopted in this work is no exception. Moreover, while the crust--core transition densities are calculated self-consistently, the random core EOSs are all matched to the same crustal EOS. In this sense, our crust and core EOSs are non-unified, which introduces an additional source of systematic uncertainty. Unfortunately, in our opinion, it is nearly impossible to quantify this uncertainty without an absolutely correct or commonly accepted crustal EOS.

Nevertheless, useful insights can be gained by comparing our results with those obtained using the latest \emph{Crust (Unified) Tool for Equation-of-state Reconstruction} (CUTER~v2) \citep{Davis:2025nwz}. For this purpose, we compare in a.Fig.~\ref{cuter} typical mass--radius sequences calculated using the NV+BPS (blue) and CUTER (black) crustal EOSs. The left (right) panel corresponds to $L=58.7$ ($40$)~MeV with all other parameters fixed as specified in the caption. As indicated by the dashed lines, the absolute difference in $R_{1.4}$ can be as large as $\sim0.083\,\mathrm{km}$, whereas it is reduced to about $\sim0.055\,\mathrm{km}$ for $R_{2.0}$ at $L=58.7$~MeV. Both differences are smaller than the highest precision of 0.1 km adopted in the mock radius data, and there likely exist other sources of systematic uncertainty of comparable or greater magnitude in our analysis.

Overall, this comparison reinforces our understanding that the radii of low-mass NSs are more sensitive to crustal physics, while the radii of massive NSs, such as $R_{2.0}$, are more suitable for probing the possible presence of quark matter cores, with minimal influence from crustal uncertainties. In future studies, it would be more physically consistent to use unified crustal EOSs, but for the work presented in the main text, using mock high-precision $R_{2.0}$ data, the use of the NV+BPS crustal EOS is physically reasonable.

\section*{\small B. Prior range and probability distribution of crust-core transition density}\label{rcc}
In our study here, while the crust EOS is fixed, the transition density $\rho_{cc}$ itself is determined self-consistently by the hadronic meta-model EOS for the outer core by setting its incompressibility $K_{\mu}$ to zero. We have studied extensively earlier in a forward-modeling approach how the uncertainties of the EOS model parameters affect the $\rho_{cc}$ \cite{Zhang:2018bwq,Zhang:2018vbw,Li:2020ass} and the resulting NS radii. Here we study how the precision of future radius measurements may better inform us about the $\rho_{cc}$ compared to its prior probability distribution function (PDF).

\begin{figure*}[ht]
\begin{center}
 \resizebox{0.9\textwidth}{!}{
  \includegraphics[width=12cm,height=4cm]{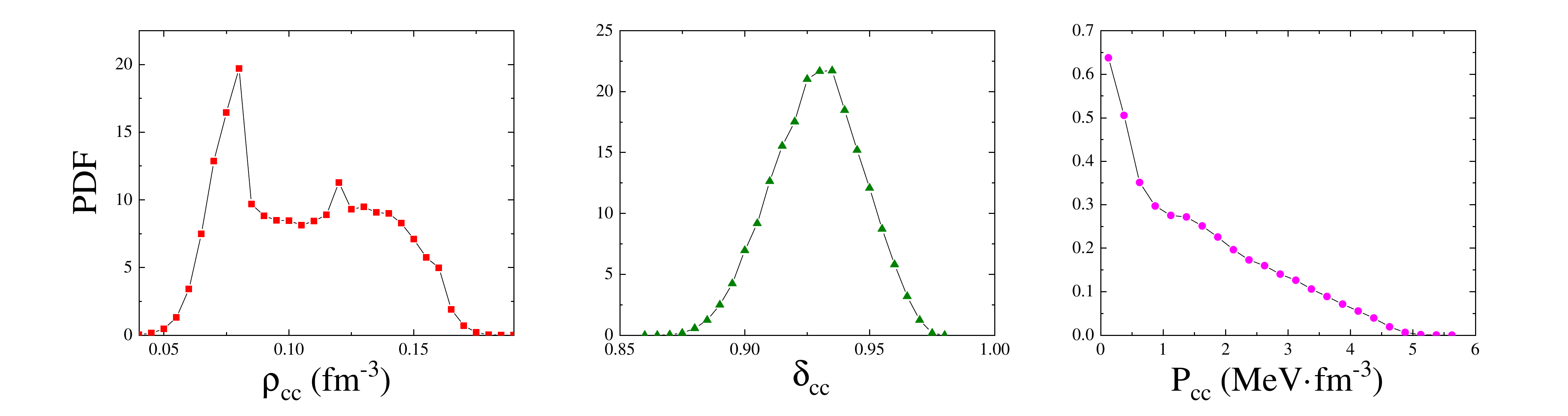}
  }
  \caption{(color online) Probability distribution function (PDF) of density $\rho_{cc}$ (left), isospin asymmetry $\delta_{cc}$ (middle) and pressure $P_{cc}$ (right) at the crust-core transition point, respectively, generated from using 40,000 EOSs for the outer core matter in neutron stars.}\label{rcc-pdf}
\end{center}
\end{figure*}

To obtain the prior PDF for $\rho_{cc}$, firstly, we prepare 40,000 meta-model EOSs for the outer core (as discussed in detail in the main text) by uniformly generating randomly the nuclear matter EOS parameters within their uncertainties listed in Table \ref{tab-prior}. 
The incompressibility $K_{\mu}$ of the outer core can be rewritten in terms of the EOS parameters as \cite{Li:2020ass}
\begin{equation}\label{kmu2}
K_{\mu}=\frac{1}{9} (\frac{\rho}{\rho_0})^2 K_0+ 2 \rho \frac{\rm{d}E_0}{\rm{d} \rho}+\delta^2
\left[\frac{1}{9} (\frac{\rho}{\rho_0})^2 K_{\rm{sym}}
+\frac{2}{3}\frac{\rho}{\rho_0}L
-2E^{-1}_{\rm sym}(\rho)(\frac{1}{3}\frac{\rho}{\rho_0}L)^2\right]. 
\end{equation}
Clearly, it depends nonlinearly on the hadronic EOS parameters in a complicated way. Thus, the root of equation $K_{\mu}=0$ is expected to be non-uniform for randomly generated EOS parameters with uniform prior PDFs in their prior ranges specified in Table \ref{tab-prior}. Shown in a.Fig. \ref{rcc-pdf} are PDFs of the resulting transition density $\rho_{cc}$ as well as the isospin asymmetry $\delta_{cc}$ and the pressure $P_{cc}$ there from the expression \cite{lattimer2007neutron}
\begin{equation}
    P_{cc}=\frac{K_0}{9}\frac{\rho_{cc}^2}{\rho_0}\left(\frac{\rho_{cc}}{\rho_0}-1\right)
    +\rho_{cc}\delta_{cc}\left[\frac{1-\delta_{cc}}{2}E_{\rm sym}(\rho_{cc})
    +\left(\rho\frac{{\rm d}E_{\rm sym}(\rho)}{{\rm d}\rho}\right)_{\rho_{cc}}\delta_{cc}\right]\nonumber.
\end{equation}
\begin{figure*}[ht]
\begin{center}
 \resizebox{0.7\textwidth}{!}{
  \includegraphics[width=12cm,height=8cm]{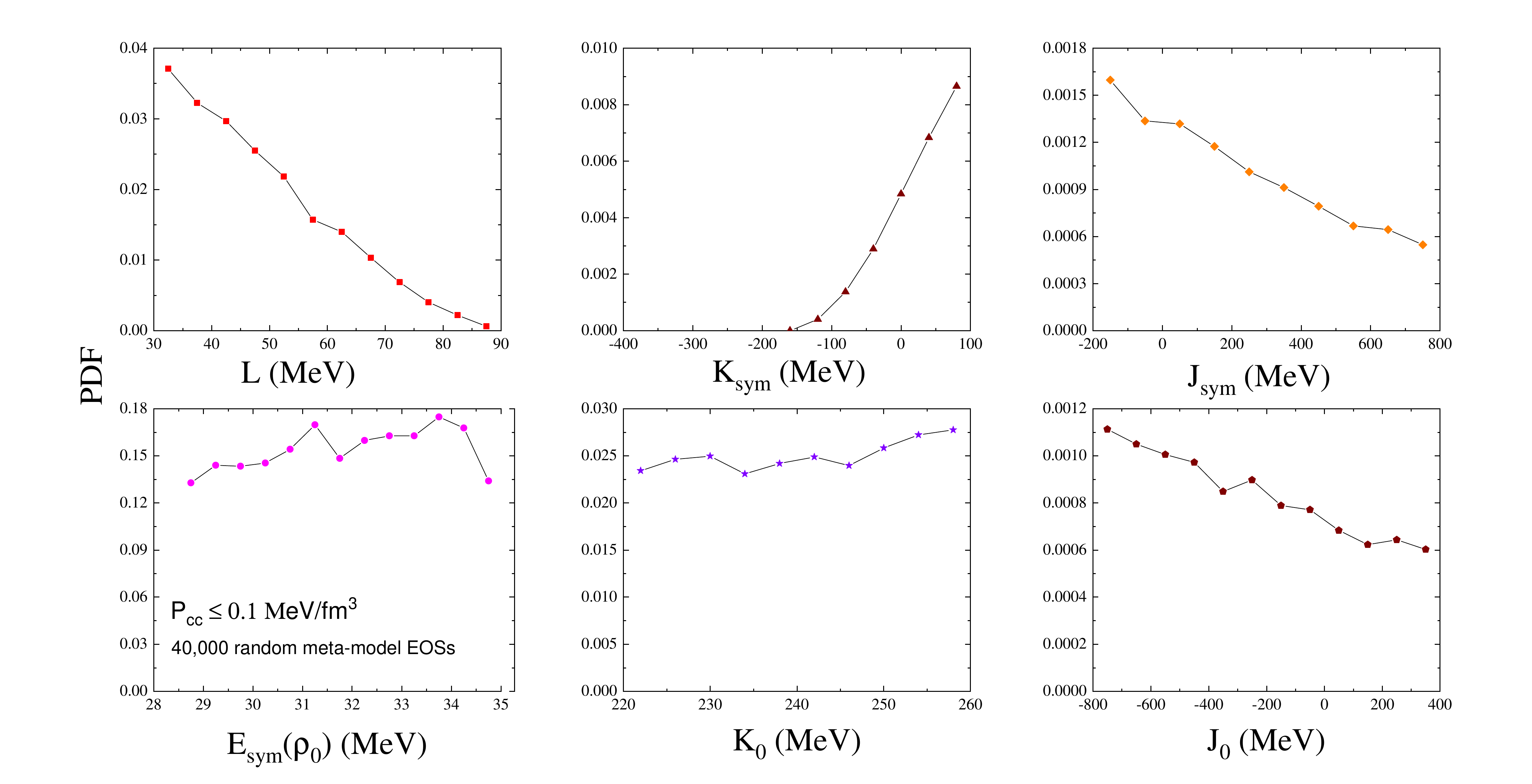}
  }
  \caption{(color online) Probability distribution functions (PDFs) of six nuclear EOS parameters leading to a pressure $P_{cc}\leq 0.1$ MeV/fm$^{3}$ at the crust-core transition point in neutron stars.}\label{eos-pdf}
\end{center}
\end{figure*}
The most probable $\rho_{cc}$ is around 0.075 fm$^{-3}$ consistent with its fiducial value of $\rho_{cc}$=0.08 fm$^{-3}$ that has long been used widely in the literature \cite{baym1971ground}. The corresponding PDFs of $\delta_{cc}$ and pressure $P_{cc}$ indicate that the transition occurs in a very neutron-rich environment at rather low pressures close to the NS surface as one expects. 

To see which hadronic EOS parameters are most important for determining the crust-core transition properties, shown in a.Fig. \ref{eos-pdf} are the PDFs of the six EOS parameters under the condition $P_{cc}\leq 0.1$ MeV/fm$^{3}$ where the PDF($P_{cc}$) peaks. Interestingly, it is seen that the PDFs of $L$, $K_{\rm{sym}}$, and $J_{\rm{sym}}$ characterizing the density dependence of nuclear symmetry energy illustrate very strong variations, indicating that they are most important for determining the crust-core transition properties. Moreover, the skewness $J_0$ of SNM through the $\rm{d}E_0/\rm{d}\rho$ term in the expression of $K_{\mu}$ also plays an appreciable role in determining $\rho_{cc}$. On the other hand, the two parameters $E_{\rm{sym}}(\rho_0)$ and $K_0$, most relevant to characterizing the properties of neutron-rich matter at $\rho_0$ have little effect. We notice that the little bumps in some of the PDFs are due to the relatively small number of EOSs and the fine bin sizes we used in preparing the illustrations here. 

\begin{figure*}[ht]
\begin{center}
 \resizebox{0.49\textwidth}{!}{
  \includegraphics[]{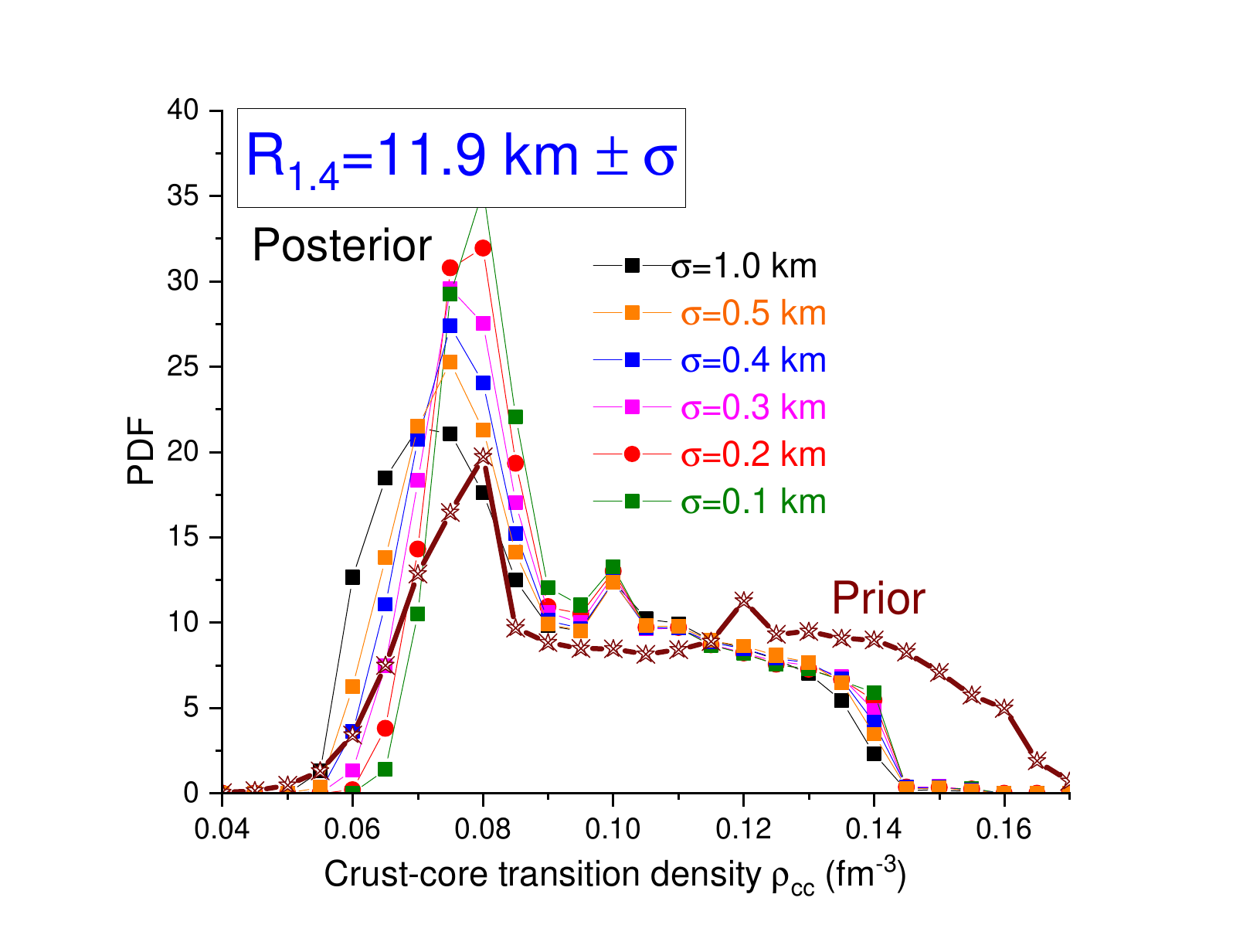}
  }
   \resizebox{0.49\textwidth}{!}{
  \includegraphics[]{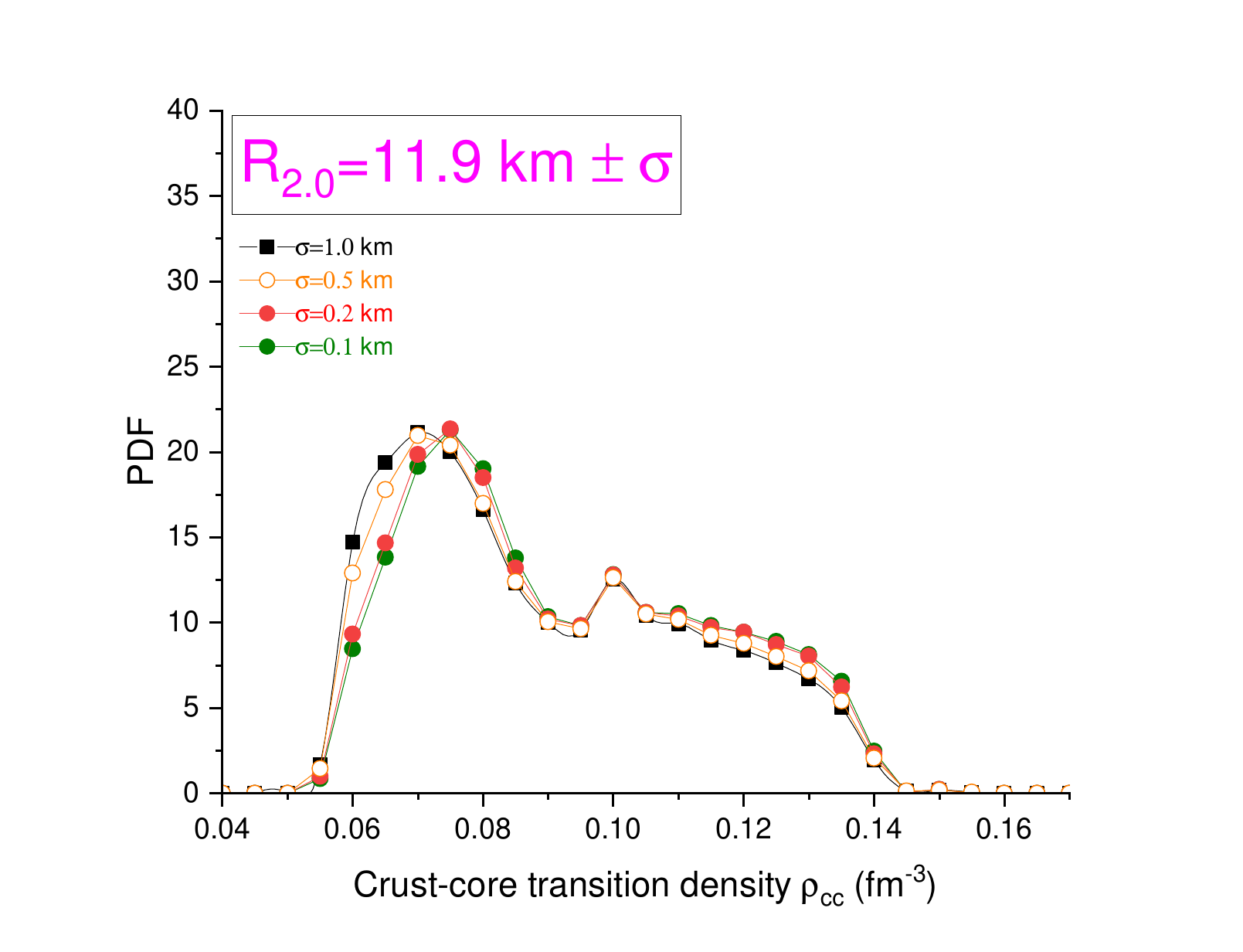}
  }
  \caption{(color online) Posterior PDFs of the crust-core transition density $\rho_{cc}$ inferred from Bayesian analyses using $R_{1.4}=11.9$ km (left where the prior PDF of $\rho_{cc}$ is also shown for a comparison) and $R_{2.0}=11.9$ km (right) 
  with varying precision $\sigma$ of future radius measurements as indicated.}\label{post-rcc}
\end{center}
\end{figure*}
\begin{figure*}[ht]
\begin{center}
 \resizebox{0.45\textwidth}{!}{
  \includegraphics[]{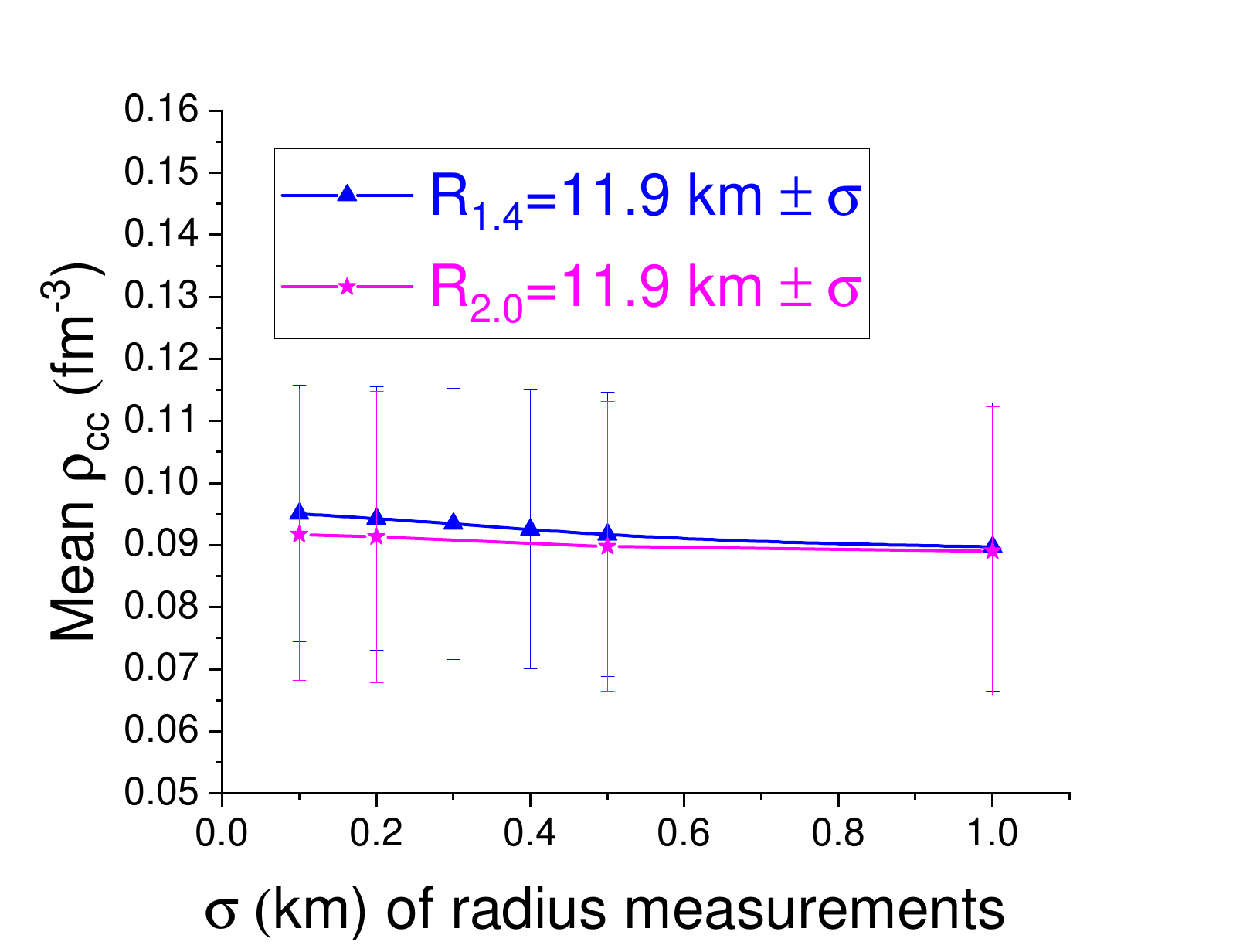}
  }
   \resizebox{0.45\textwidth}{!}{
  \includegraphics[]{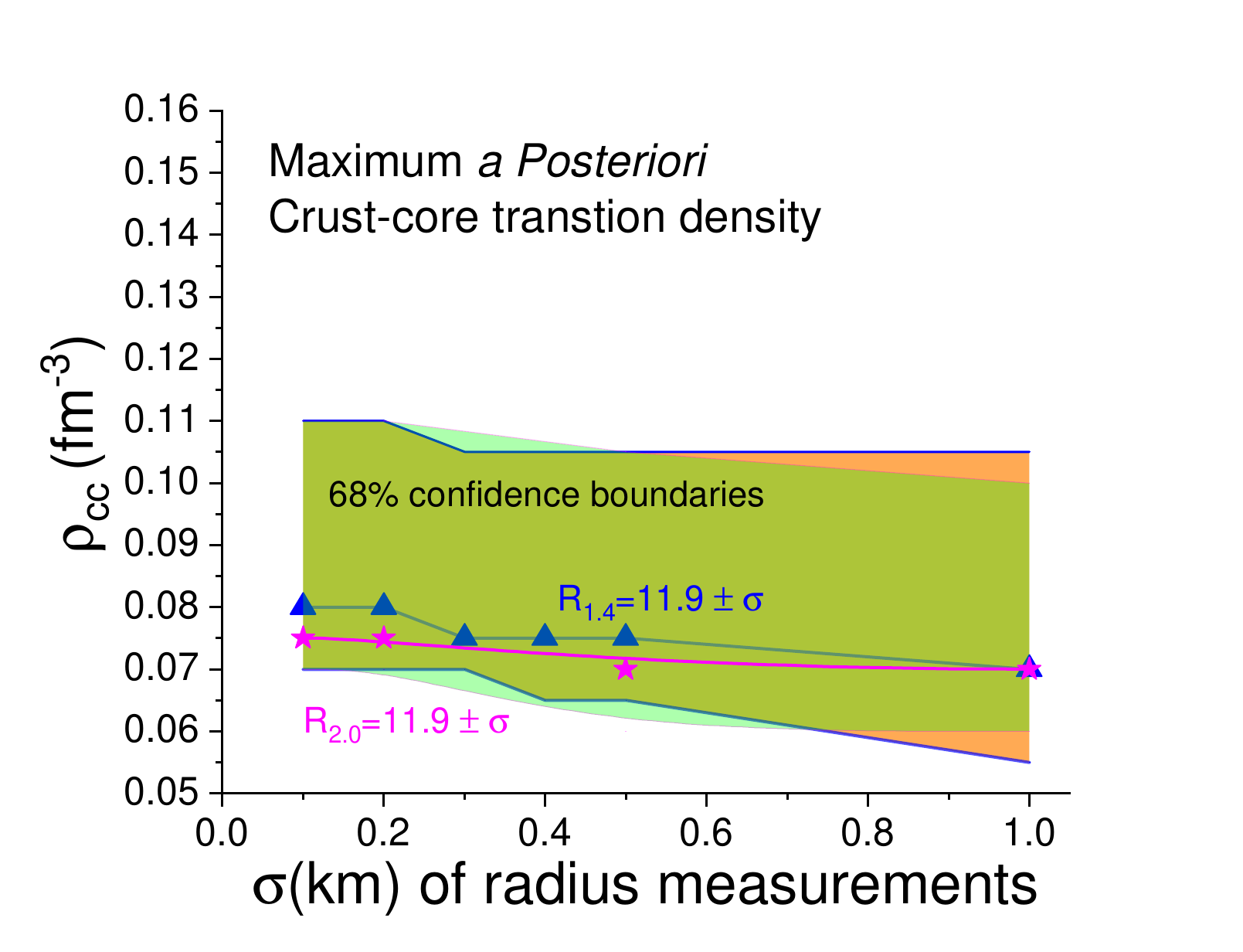}
  }
  \caption{(color online) The mean and standard deviation of $\rho_{cc}$ (left) as well as the 
Maximum {\it a Posteriori} (MaP) and its 68\% confidence boundaries of the posterior PDF($\rho_{cc}$) (right) as functions of the precision $\sigma$ in measuring NS radii.}\label{mean-max}
\end{center}
\end{figure*}

\section*{\small C. Posterior PDF of crust-core transition density from future high-precision NS radius measurements}\label{rcc2}
\indent
Would future high-precision NS radius measurements help narrow down the uncertainty range of the crust-core transition density? In other words, would the uncertainties associated with the low-density hadronic EOS in determining the $\rho_{cc}$ prevent us from learning new physics about the high-density NS inner core using high-precision radius measurements? To answer these questions, shown in a.Fig.~\ref{post-rcc} are posterior PDFs of the crust-core transition density $\rho_{cc}$ inferred from our Bayesian analyses using $R_{1.4}=11.9$ km (left) and $R_{2.0}=11.9$ km (right) with varying precision $\sigma$ expected for future radius measurements as indicated. The prior PDF of $\rho_{cc}$ discussed above is also shown in the left panel for a comparison. We emphasize that the posterior PDF($\rho_{cc}$) presented here is obtained from the NS minimum model without considering the possible hadron-quark phase transition to avoid effects caused by the uncertainties associated with the latter. As shown in detail in the main text, the uncertain prior ranges of the three quark matter EOS parameters affect somewhat the posterior PDFs of the six hadronic EOS parameters inferred from NS radius data. They may thus have some small secondary effects on the posterior PDF($\rho_{cc}$) if we use the full EOS model encapsulating a hadron-quark phase transition. Moreover, as our main goal is to study how high-precision NS radius measurements may help improve our knowledge about the high-density quark matter, comparing the prior and posterior PDFs of $\rho_{cc}$ both obtained without considering the quark core provides a more meaningful and clean reference. Several interesting observations can be made qualitatively from inspecting the results shown in a.Fig. \ref{post-rcc}. In particular,
\begin{itemize}
    \item The upper boundary of $\rho_{cc}$ in its posterior PDF is somewhat lower than that in its prior PDF ( shifted from about 0.17 in its prior to 0.14 fm$^{-3}$ in its posterior PDF), illustrating clearly the visible power of NS observational data in constraining the crust-core transition density.
    \item With $R_{1.4}=11.9$ km, as the precision improves from $\sigma=1.0$ to 0.1 km, the most probable value of $\rho_{cc}$ apparently shifts towards and saturates narrowly around the fiducial value of $\rho_{cc}\approx 0.08$ fm$^{-3}$, indicating that the radii of canonical NSs are more affected by the uncertainties associated with the crust-core transition. 
    \item With $R_{2.0}=11.9$ km, the posterior PDF of $\rho_{cc}$ clearly has less dependence on the precision $\sigma$, and it peaks around $\rho_{cc}=0.075$ fm$^{-3}$ slightly lower than that with $R_{1.4}=11.9$ km, indicating that the radii of massive NSs are not much affected by the uncertainties associated with the crust-core transition.
\end{itemize}

To be more quantitative, shown in a.Fig. \ref{mean-max} are the mean and standard deviation of $\rho_{cc}$ (left) as well as the MaP and its 68\% confidence boundaries of the posterior PDF($\rho_{cc}$) (right) as functions of the precision $\sigma$ in measuring NS radii. We notice that the MaP values are closer to the lower boundary of the 68\% confidence interval because the highly-asymmetric PDFs of $\rho_{cc}$ have long tails towards higher densities. It is seen that both the mean and the MaP values inferred from using $R_{1.4}=11.9$ km or $R_{2.0}=11.9$ km are approximately the same, with essentially the best precision of $\sigma=1.0$ km currently available. As the precision improves, while both the mean and the MaP values inferred from using $R_{2.0}=11.9$ km remain approximately the same, the results from using $R_{1.4}=11.9$ km clearly become slightly larger.

\bibliographystyle{plainnat}
\bibliographystyle{aasjournal}
\bibliography{references}

\clearpage
\end{document}